\documentclass[a4paper,11pt]{article}
\pdfoutput=1 % if your are submitting a pdflatex (i.e. if you have
\usepackage{jcappub} % for details on the use of the package, please
\usepackage{natbib}

\usepackage[T1]{fontenc} % if needed
\title{\boldmath A Template-based $\gamma$-ray Reconstruction Method for  Air Shower Arrays}

\author[a,1]{Vikas Joshi,\note{Corresponding author.}}
\author[a]{Jim Hinton,}
\author[a]{Harm Schoorlemmer,}
\author[a,b]{Rub\'{e}n L\'{o}pez-Coto,}
\author[a]{Robert Parsons}
\affiliation[a]{Max-Planck-Institut f\"ur Kernphysik, P.O. Box 103980, D 69029 Heidelberg, Germany}
\affiliation[b]{now at Universit\`a di Padova and INFN, I-35131, Padova, Italy}
\emailAdd{vikas.joshi@mpi-hd.mpg.de}
\abstract{We introduce a new Monte Carlo template-based reconstruction method for air shower arrays, with a focus on shower core and energy reconstruction of $\gamma$-ray induced air showers. The algorithm fits an observed lateral amplitude distribution of an extensive air shower against an expected probability distribution using a likelihood approach. A full Monte Carlo air shower simulation in combination with the detector simulation is used to generate the expected probability distributions. The goodness of fit can be used to discriminate between $\gamma$-ray and hadron induced air showers. As an example, we apply this method to the High Altitude Water Cherenkov $\gamma$-ray Observatory and its recently installed high-energy upgrade. The performance of this method and the applicability to air shower arrays with mixed detector types makes it a promising reconstruction approach for current and future instruments.}

\begin{document}
\maketitle
\flushbottom
\section{Introduction}
\label{intro}
Extensive Air Shower (EAS) detector arrays take advantage of their large collection area on the ground to detect the secondary particles generated by the interaction of a primary particle in the Earth's atmosphere. The estimation of EAS properties is performed by measuring the Lateral Distribution Function (LDF) and the arrival time distribution of the secondary particles. In this paper, we demonstrate a method to estimate properties of the EAS that can be derived from the information in LDF. The LDF of an EAS describes the observed number of particles at a given distance from the shower axis (impact distance). By using the LDF information the shower impact point on the ground and energy of the primary particle can be estimated.

Traditionally, the LDF is fitted with a functional shape, which is typically derived empirically to describe on average the features of the distribution. Typically, the parameters of these functions cannot directly be associated with the air shower properties. A widely used functional shape is the Nishimura-Kamata-Greisen (NKG) functions \citep{ref_NKG} \citep{ref_cosmic_ray_showers}. In addition to the shape of the distribution, also the fluctuations needed to be taken into account in the fitting procedure. Fluctuations in observed signal amplitude arise from detector response and fluctuations in particle densities, both of which might not be trivial to parametrize. This makes the fit results from these approaches to depend on the actual type of the detector, in turn making the combined fitting of EASs with a mixed array of particle detectors challenging. 

To address these problems, we present a Monte Carlo (MC) template-based likelihood fit method for $\gamma$-ray induced EASs observed with an array of particle detectors. This method can in principle be extended to air showers induced by other cosmic-ray particles, however, our focus is on the application of this method in $\gamma$-ray astronomy. 
The template-based fit procedure for $\gamma$-ray induced EAS was pioneered for the CAT telescope \citep{ref_CAT1} \citep{ref_CAT2} and further improved and re-implemented in the H.E.S.S telescopes \citep{ref_CAT_method_in_HESS}. 
A more mature version of this approach based on MC templates known as Image Pixel-wise fit for Atmospheric Cherenkov Telescopes (ImPACT) \citep{ref_IMPACT} is used in Imaging Atmospheric Cherenkov Telescopes (IACTs) such as H.E.S.S. and has shown its effectiveness \citep{ref_HESS}. Another example of template-based likelihood fitting of EAS was recently developed to obtain cosmic-ray energy for observations made with the High Altitude Water Cherenkov (HAWC) $\gamma$-ray Observatory. It has been applied to obtain the all-particle cosmic-ray energy spectrum recently published in \citep{hawc_all_particle_spectrum}. 

The MC template-based likelihood fit method presented here is applicable for EAS detector arrays and reconstructs the shower core and the energy of VHE $\gamma$-ray showers. In addition, the quality of the fit can be used to discriminate between EAS induced by $\gamma$-rays or hadronic particles. The nature of likelihood fitting makes it straightforward to combine measurements of different detector types in the same fit algorithm. The MC template-based likelihood fit method employs no approximations apart from the EAS simulation and the detector model itself, which is inevitable in a model-based fitting procedure. Due to the lack of hadronic interaction, $\gamma$-ray induced EASs can be modelled more reliably than nuclei. In addition, a prior unknown composition of the flux of subatomic nuclei makes the phase-space to be covered by the EAS templates significantly larger and therefore the fitting procedure more complex. 

This approach can be used in general for air shower arrays. As a proof of the concept, we demonstrate its usability for  HAWC $\gamma$-ray  Observatory \citep{ref_hawc} and to its high-energy upgrade consisting  of a sparse array of smaller water Cherenkov detectors \citep{hawc_Outriggers2}. It is applicable to the current (Tibet AS-$\gamma$ Experiment \citep{Tibet_array} and  ARGO-YBJ \citep{ARGO}) and future observatories (LHAASO \citep{LHAASO} and a next-generation observatory to be built in the Southern Hemisphere \citep{ALPACA, ALTO, LATTES,STACEX,MPI}). A method like this one is a robust approach to perform the combined reconstruction for mixed type particle detector arrays (in this case HAWC and its high energy upgrade with a sparse array).

\section{General Considerations} 
\label{considerations}
Before going into details on the method, we discuss in this section the general considerations that lead to our particular implementation. As an illustration, we use a simulated array of water-Cherenkov detectors, of which the details are given and Section \ref{simulation dataset}. However, the discussion in this section applies to all type of particle detectors.

\begin{figure}[!h]
\centering
\includegraphics[width=0.7\linewidth]{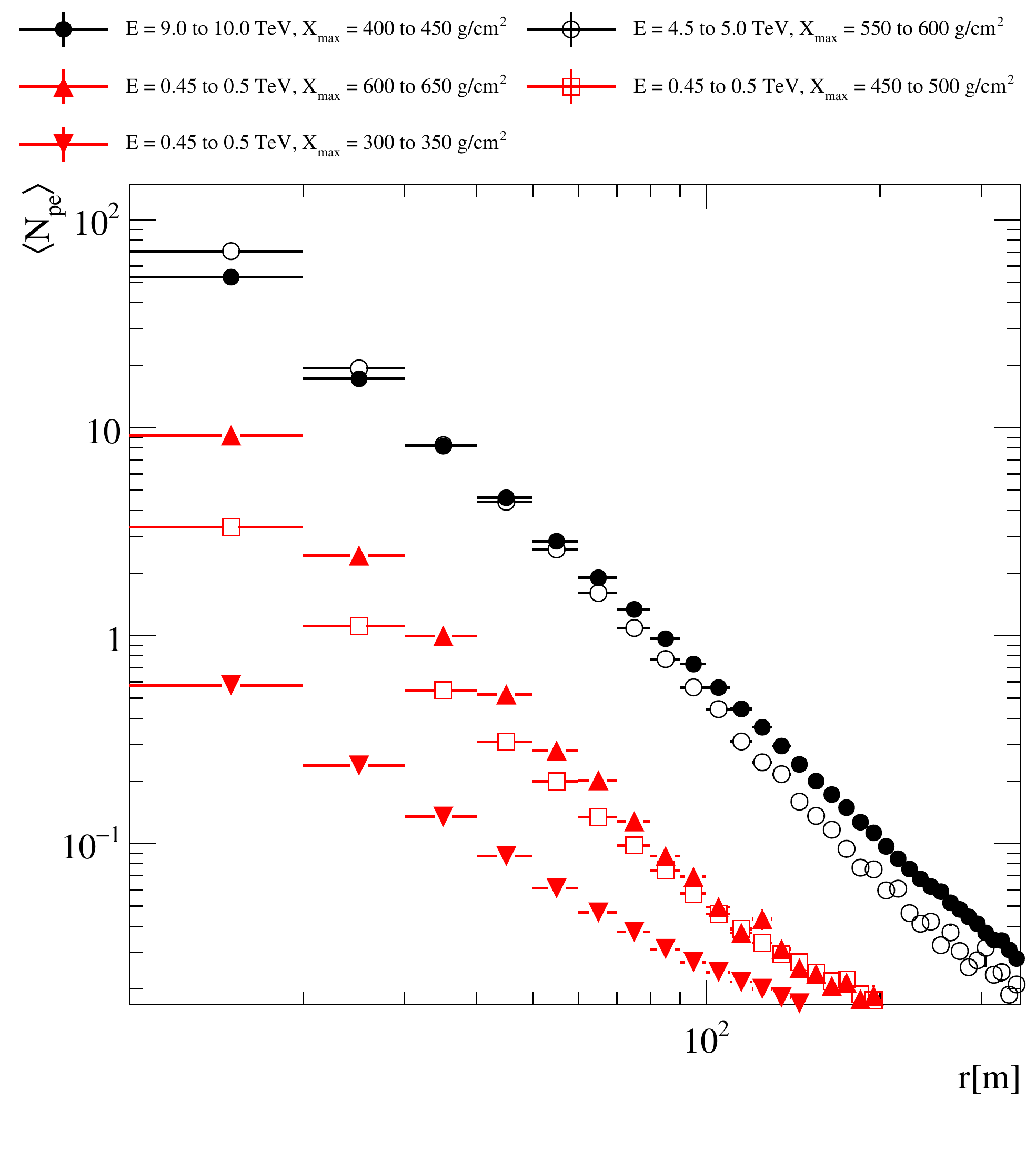}
\caption{Average of observed number of photo-electrons (N$_{\rm pe}$) as a function of the impact distance (r).}
\label{fig:energy_and_xmax_ambiguity}
\end{figure}

\subsection{Ambiguity Between Primary Energy and  X$_{\rm {max}}$}
\label{ambiguity in Energy and Xmax}
The LDFs measured by EAS detector arrays exhibit large fluctuations, due to the fact that they are observed at a particular developmental stage of the EAS. The intrinsic uncertainty in the first interaction of the primary particle gives rise to different observed LDFs for the same primary particle properties. As we are using water-Cherenkov detectors for illustration throughout this article, the LDFs are measured using the observed number of photo-electrons (pe) (for more details see Section \ref{simulation dataset}). In Figure \ref{fig:energy_and_xmax_ambiguity} averaged LDFs for $\gamma$-ray induced EAS are shown for vertical showers. The averages are calculated over EASs initiated with the same energy of the primary $\gamma$-rays that have similar shower development.  We show these curves to illustrated some complications that arise when using LDFs to deduce the properties of the primary particle.  The black curves show the example of matching LDFs combination for multi-TeV $\gamma$-rays, in which, the energies are two-fold apart and the difference in X$_{\rm max}$ values is 150 g/cm$^2$.  The red curves show that for sub-TeV $\gamma$-rays for a fixed value of energy of the primary particle with increasing X$_{\rm max}$ the averaged LDFs of the showers shifts significantly to the higher pe values, which again might lead to ambiguity in determining the primary particle energy. It is not likely to break this ambiguity completely, however by guiding the fit procedure the ambiguity can be significantly reduced. The way we implemented this will be discussed in Section~\ref{likelihood fitting}.

\subsection{Signal Free Detectors}
\begin{figure*}[!h]
\centering
\includegraphics[width=\linewidth]{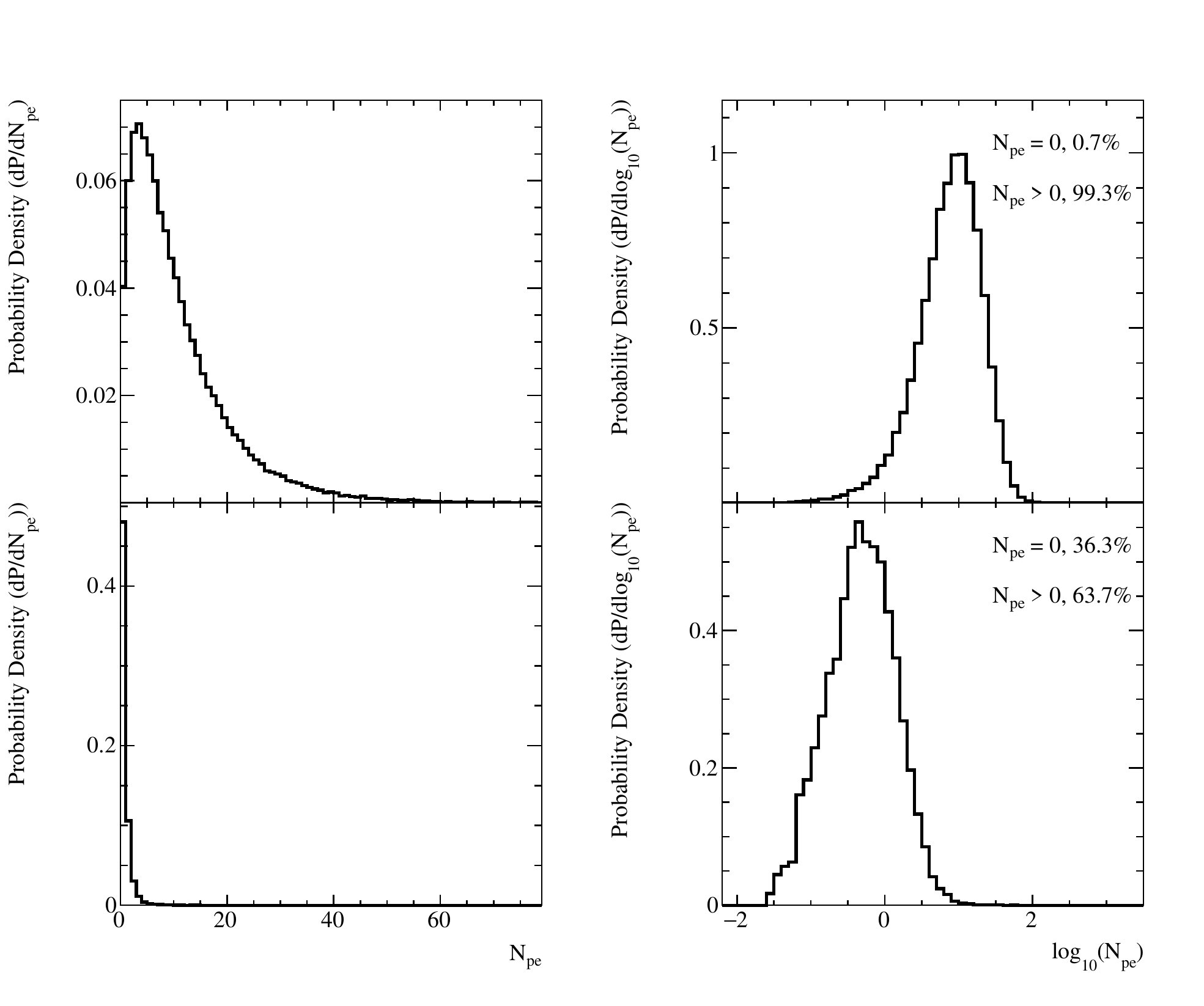}
\label{fig:energy_effect_on_logprob_profiles}
\caption{This plot includes vertically simulated EASs between 4.5 to 5.5 TeV. The top left panel shows the probability density  distribution of the observed number of pe (N$_{\rm pe}$) at an impact distance of 20 m and the right panel shows it for log$_{10}$(N$_{\rm pe}$). The bottom panel (left and right) show the same as above at an impact distance of 80 m.}
\label{fig:fluctuations_in_obsereved_number_of_pe}
\end{figure*}

The LDF has a long tail in which typically only a fraction of the detectors will have a signal, in Figure  \ref{fig:energy_and_xmax_ambiguity} this corresponds to the distances at which the value of $\langle {\rm N_{pe}}\rangle < 1$~pe. It can be seen that there actually is some information in the tail of distribution that can break the aforementioned ambiguity, therefore it is desirable to maximise the information in the fit which includes the measurement of zeros. In Figure \ref{fig:fluctuations_in_obsereved_number_of_pe} we show the probability density distributions of observing a certain number of pe at impact distances of $\sim$20 m and $\sim$80 m. We see that the probability of observing zero pe signal increases from 0.7\% to 36.3\% at impact distances of $\sim$20 m to $\sim$80 m. In a likelihood-based method, it is straightforward to add the LDF of the signal free detectors to the likelihood function. 

\subsection{Logarithmic Signal Amplitude}
Since the underlying physical processes that lead to the particle densities observed at the ground are partially multiplicative in nature, fluctuations are typically easier described in log-space. This is illustrated in Figure \ref{fig:fluctuations_in_obsereved_number_of_pe} in which the left panels show the amplitude distribution in a linear scale, while the figure on the right shows them on log-scale. It can be observed, by  transforming to log-space the effect of the long-tails in the amplitude are significantly reduced. And in addition, small signal distributions can be described with similar binning as that of large signals, making the generation of template distributions for the likelihood fit significantly easier in log-space. The observed zero pe signal is stored at a low negative value in the log-space, which is lower than the small non-zero pe signal observed in the log-space.

\section{Likelihood Function}
\label{likelihood function}
Due to the aforementioned reasons, we give an alternative to a semi-analytical model dependent fit and present an MC template-based likelihood fit method. It automatically takes into account all the fluctuations and gives a complete picture of the model in a probabilistic way with only the assumption that the MC model and the detector simulations are accurate enough. 

In Figure \ref{fig:lookuptable_before_and_after_smoothing}, we show one of the MC templates which describes the probability of an observed LDF of a $\gamma$-ray shower for given shower parameters. 
These templates are generated by binning the simulated MC dataset in  Energy (E), X$_{\rm max}$, and zenith angle ($\theta$) bins. One such three-dimensional bin contains one such template, further binned in the logarithm of the observed number of pe ($\log_{10}({\rm N}_{\rm pe}$)) and the impact distance of the pe signal (r) bins. However, it is to be noted that the observed signal does not necessarily have to be observed pe. For different particle detection techniques, the nature of the observed signal may vary. The function to be minimised in the fit procedure is
defined as the negative log-likelihood

\begin{equation}
\log L = - 2 \sum_i \log(F(S_i,r_i,X_{\rm max},E | \theta,\phi)),
\label{eqn: likelihood eq}
\end{equation}

where function $F$ gives the probability of a given detector unit observing signal ($S_i$) situated at an impact distance of ($r_i$) for a $\gamma$-ray shower of energy ($E$), ($X_{\rm max}$),  zenith angle ($\theta$) and azimuth angle ($\phi$). For illustration purpose, in this article for water-Cherenkov detector technique, $S_i$ term is replaced with $\log_{10}({\rm N}_{\rm pe})_i$. Distance $r_i$  is defined as the perpendicular distance to the shower axis 
\begin{equation}
r_i = [(x_i - x_{\text{c}})^2+(y_i - y_{\text{c}})^2
-\sin^2\theta\{(x_i-x_{\text{c}})\cos\phi+(y_i-y_{\text{c}})\sin\phi\}^2]^{1/2},
\label{eqn:dist shower plane eq}
\end{equation}
with $x_i$, $y_i$ and $x_{\text{c}}$, $y_{\text{c}}$ representing the coordinates of the different detector units and the location of the shower core in the detector plane respectively. The calculation of $r_i$ in shower frame, as shown in equation \ref{eqn:dist shower plane eq} contains the information of the core coordinates on the ground. 

For a given EAS, one can get the log-probability for all detector units using the previously defined templates. Summing-up the log-probability of all detector units gives the likelihood of the given set of values of the parameters. As we are using negative log-likelihood, by minimising the likelihood ($L$) one obtains the best fitting values at the minimum value of $L$, the value of $x_{\text{c}}$ and $y_{\text{c}}$ give the estimated shower core coordinates,  $E$ gives the estimated primary $\gamma$-ray energy, and $X_{\rm max}$ gives the depth of the shower maximum. 

It is straightforward to extend this formalism for a mixed type particle detector array. One can define the likelihood function as defined in equation (\ref{eqn: likelihood eq}) for the different type of detector arrays and then summing them will give the total likelihood function (see equation \ref{eqn:total likelihood}) for the combined detector array
\begin{equation}
L_{\rm total} = L_{\rm type_1} + L_{\rm type_2} + L_{\rm type_3} + ...,
\label{eqn:total likelihood}
\end{equation}
where $L_{\rm type_i}$ tells the likelihood of detector type $i$ and it can be described as in equation (\ref{eqn: likelihood eq}) with the corresponding probability function $F_\text{type}$.

To assess the quality of the likelihood fit, we construct a Goodness of Fit (GoF). The GoF value can be used to identify events on which the  fit method failed. Further, it can also be used to deliberately separate the model behaviours which are different than the model for the fit method itself. In this case, the model is defined for the $\gamma$-ray induced air showers, and the other models are the LDFs of a hadron induced air showers.  We define our GoF by modelling the mean and the uncertainty of the likelihood value as a function of observed percentage hits of the array. For a given value of the percentage hits of the array, the resultant distribution for the likelihood values results in a Gaussian-like distribution. Therefore, we fit a Gaussian to the likelihood distribution for a given value of the percentage hits of the array, the mean $\langle L\rangle$ and $\sigma$ of the Gaussian are used to define the GoF in the following way
\begin{equation}
\text{GoF} = \frac{L_{\rm Fit} - \langle L\rangle}{\sigma}.
\label{eqn: Goodness of fit}
\end{equation}

\section{Air Shower and Detector Simulations}
\label{simulation dataset}
Detailed MC simulations are needed in order to generate the MC based templates. 
In order to simulate the interactions in the atmosphere induced by the $\gamma$-ray, we use the CORSIKA package (v7.4000) \citep{ref_CORSIKA}, which provides us with secondary particles tracks at the ground level. To obtain the LDF-templates, we used a large statistic set of simulations in the energy range between 0.3 and 300 TeV and with an $E^{-2}$ energy spectrum. The zenith angles ($\theta$) of these $\gamma$-rays were distributed uniformly in $\cos \theta$ within the range $ 0^\circ < \theta< 45^\circ$. 

Throughout this article, we will show the method applied to the HAWC $\gamma$-ray observatory. HAWC is situated at the Sierra Negra in Mexico at an altitude of 4100 m a.s.l. It consists of 300 cylindrically shaped Water Cherenkov Detectors (WCDs) of 7.3 m diameter and 5 m in height. Each of these WCDs has three 8", and one 10" upward facing Photo Multiplier Tubes (PMTs) anchored at the bottom that detect the Cherenkov light produced by secondary particles passing through the water in the tank. The LDF of an EAS is obtained from the charge, expressed in the number of pe, observed at each PMT.  The detector response, including interactions of secondary  particles in the WCDs, Cherenkov light production, propagation, and detection by PMTs are modeled using a dedicated software package based on Geant4 (v4.10.00) (HAWCSim) \citep{ref_GEANT4}. A more detailed description of the HAWC simulations can be found in  \citep{ref_HAWC_crab_paper}.

\section{Template Generation}
\label{template generation}
In this section, we explain the different steps in the procedure to generate MC based templates using the simulation dataset defined in Section \ref{simulation dataset}.

\subsection{Binning Scheme}
\label{binning scheme}

We have binned the dataset described in Section \ref{simulation dataset} in  $E$, $X_{\rm max}$ and $\theta$ bins. The binning of the parameter phase-space has been optimised as a compromise between the size of the dataset and the achievable resolution on the fit parameters. The optimisation procedure is described in section \ref{Optimisation of Binning}, here we summarise the resulting binning scheme:

\begin{itemize}
\item 5 $\theta$ bins (0 to $45^{\circ}$): equally spaced in $\cos\theta$, bin size = 0.06
\item  30 $E$ bins (0.3 to 300 TeV): equally spaced in $\log_{10}$(E/\text{GeV}), bin size = 0.1
\item 12 $X_{\rm max}$ bins (150 to 750 g/cm$^2$): bin size = 50 g/cm$^2$
\end{itemize}
Each combination of $E$, $X_{\rm max}$ and $\theta$ bin contains a PDF as shown in Figure \ref{fig:lookuptable_before_and_after_smoothing} for the probabilistic description of an observed lateral amplitude distribution of a $\gamma$-ray shower. We stored the PDFs as two-dimensional (2D) histograms with $\log_{10}$(N$_{\rm pe}$) on one axis and $r$ on the other. The binning inside these 2D histograms is defined as follows:
\begin{itemize}
\item $\log_{10}$(N$_{\rm pe}$) bins: bin size = 0.1 (which is approximately equal to the charge resolution of PMTs in use ~33\%)
\item $r$ bins: bin size = 2 m, in range of 0 to 500 m.
\end{itemize} 

\subsection{Smoothing}
\label{smoothing}
Since the MC simulation chain is  very computationally intensive, generating enough MC statistics to populate the whole phase-space of EAS parameters is not practical. Therefore, there are very rarely or unpopulated bins that occur particularly in the edges of the phase-space which contain little statistics. These bins might introduce unwanted artefacts in the templates that might influence the fits. 

\begin{figure*}[!h]
\centering
\includegraphics[width=0.495\linewidth]{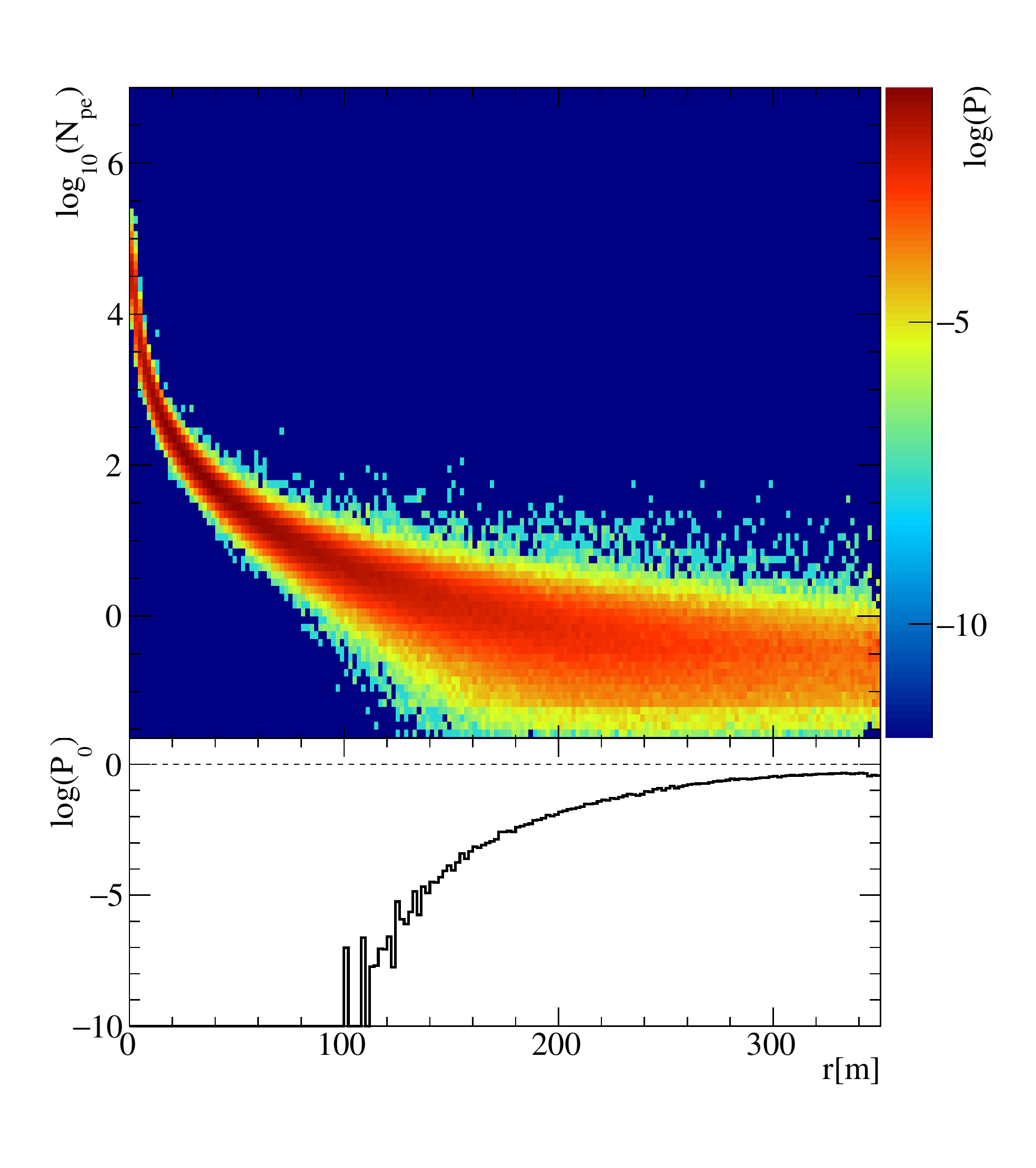}
\includegraphics[width=0.495\linewidth]{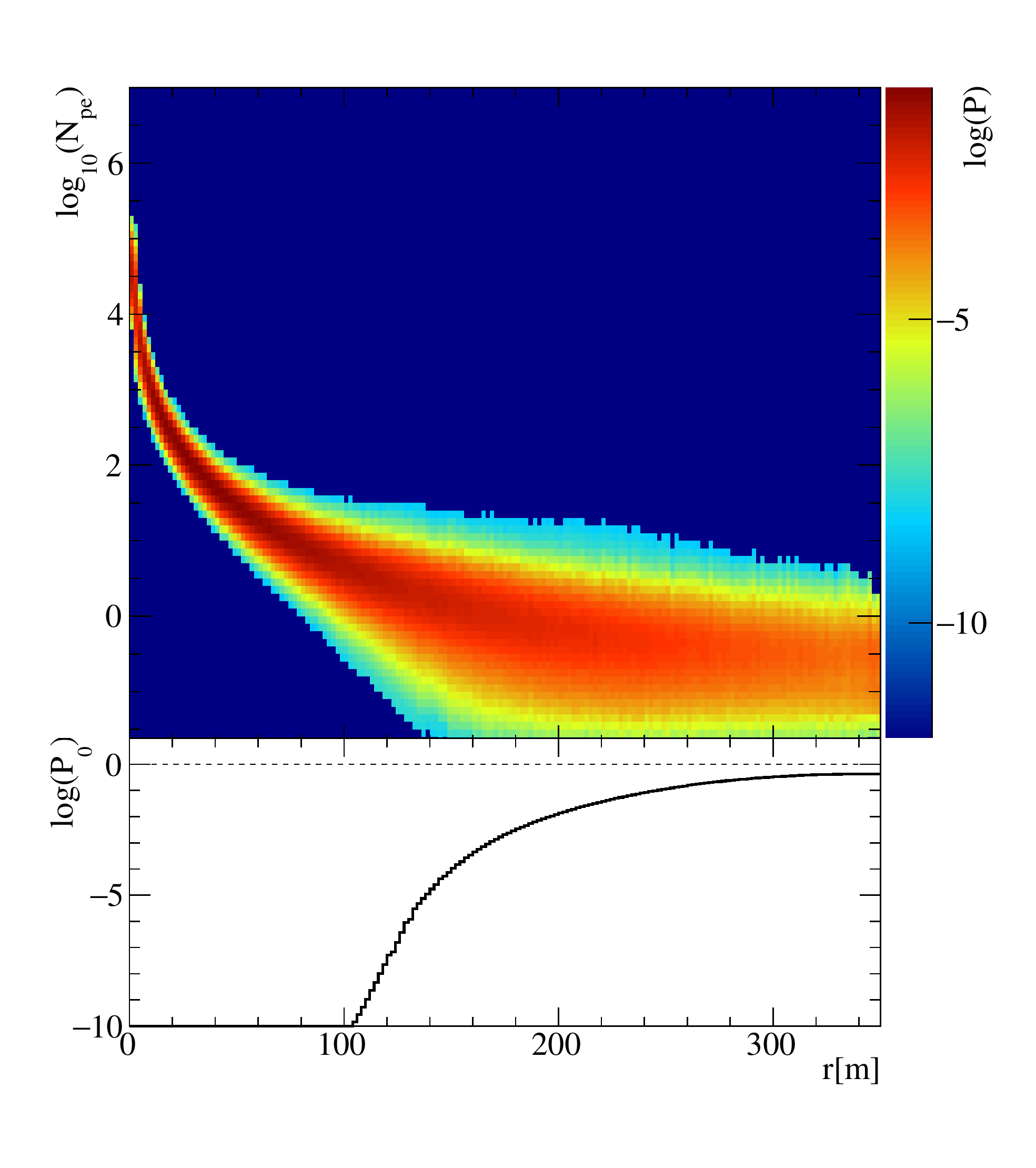}
\caption{A typical template histogram for X$_{\rm max}$ bin 500 to 550 g/cm$^2$, energy bin $\sim$59 to $\sim$75 TeV and zenith angle bin 19 to 28 degree. It shows the probability (P) distribution as a function of $\log_{10}$(N$_{\rm pe}$) and impact distance (r). The z-axis shows the probability in logarithmic scale. The left panel shows the template histogram before smoothing and the right one after smoothing. The histograms attached below show the probability (P$_{0}$) of observing a zero pe signal as a function of $r$.}
\label{fig:lookuptable_before_and_after_smoothing}
\end{figure*}

To reduce the impact of these bins, we smoothed the PDFs using Gaussian distributed weighted sum for a given bin in a given direction ($r$ or log$_{10}$(N$_{\rm pe}$)). Figure \ref{fig:lookuptable_before_and_after_smoothing} shows one such profile before and after smoothing.

\subsection{Interpolation}
\label{interpolation}
It is essential to make the likelihood surface smooth specifically for a multidimensional fitting procedure such as this case. We used three-dimensional grid interpolation over the parameters (i.e.  $E$, $X_{\rm max}$ and $r$) to obtain the probability of a given signal at a given impact distance. This probability is then used in the likelihood fit procedure to calculate the likelihood value.

\section{Fitting Procedure}
\label{likelihood fitting}
As described in Section \ref{likelihood function}, we get our likelihood function by combining the probabilities of observing $\log_{10}$(N$_{\rm pe}$) at a given distance $r$ for given values of  $E$, $X_{\rm max}$ and $\theta$ of the primary particle at different detector units. In this section, we will further describe the details of the likelihood fit method itself explaining the various measures we took to aid the fit method to find the global minimum. The fitting is performed using the widely-used MINUIT \citep{ref_MINUIT} minimisation algorithm.

We have four parameters  $E$, $X_{\rm max}$, $x_{\text{c}}$ and $y_{\text{c}}$ to fit simultaneously for a given information of shower arrival direction. In order to guide the minimisation process, we applied the strategy described below. The reconstruction of the direction is not part of the method presented in this paper, however, it is needed to select the right lookup table and to transform the detector locations into the plane perpendicular to shower axis. For direction reconstruction, we used a curved shower front fit which is used within the HAWC software framework as input for our method. The reconstructed direction together with the calculated Centre-Of-Mass (COM) of the amplitude of the signals provides the starting point of the fit procedure.

\subsection{First Pass}
 In the first pass of our method, we try three starting values for the minimisation procedure, based on the calculated COM.  
First being the COM estimate itself, which in the case of a shower core inside the array should provide a starting point close the global minimum. In order not to get stuck in a local minimum, we start the minimisation procedure at two additional starting ($X$,$Y$) locations. These trials are defined in the direction of the COM estimate with respect to the centre of the array, which is an educated guess of the true location of the shower core. The second trail is just outside the array, while another one is placed at a significant distance from the array. 

\begin{figure*}[!h]
\centering
\includegraphics[width=0.495\linewidth]{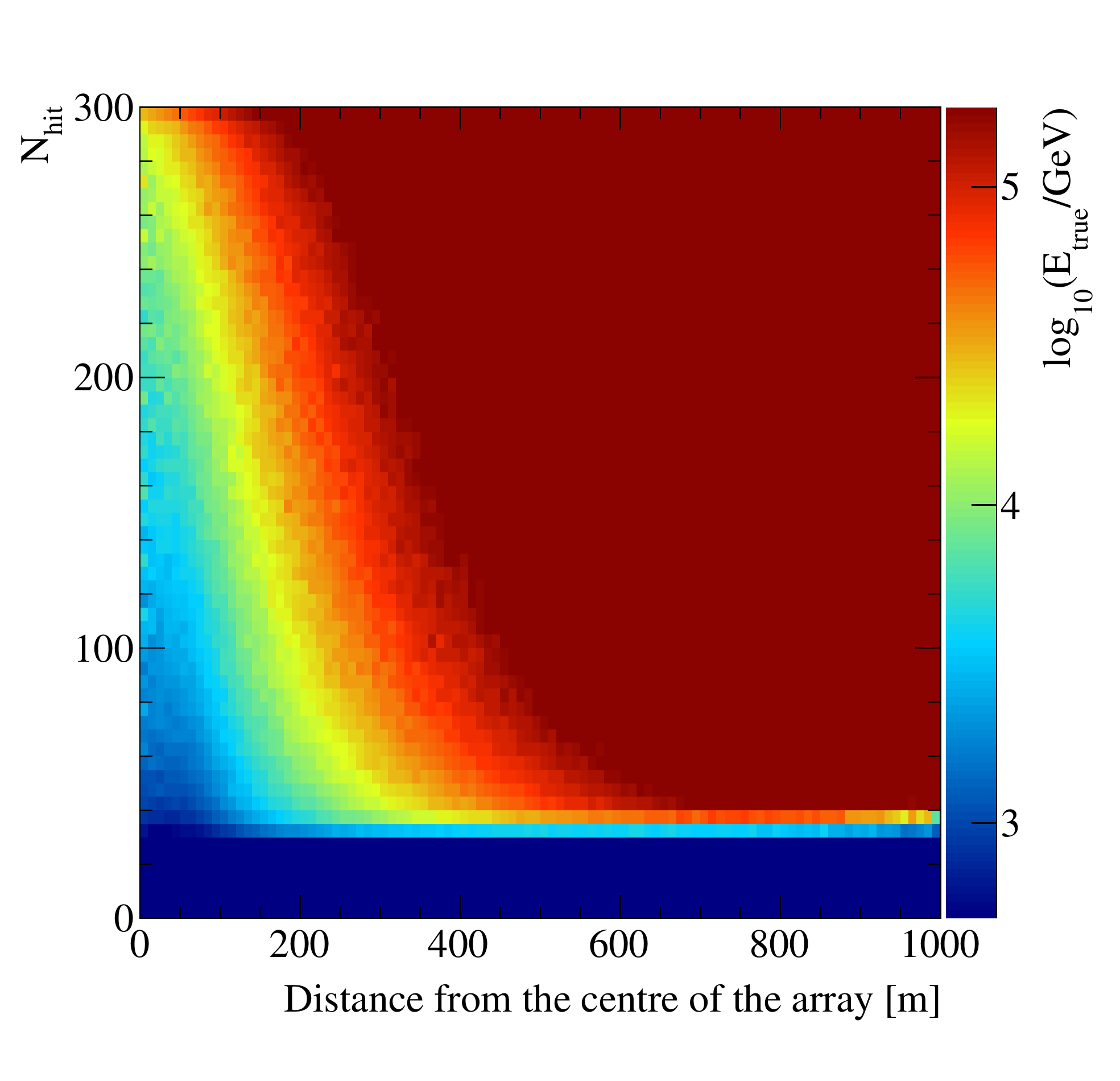}
\includegraphics[width=0.495\linewidth]{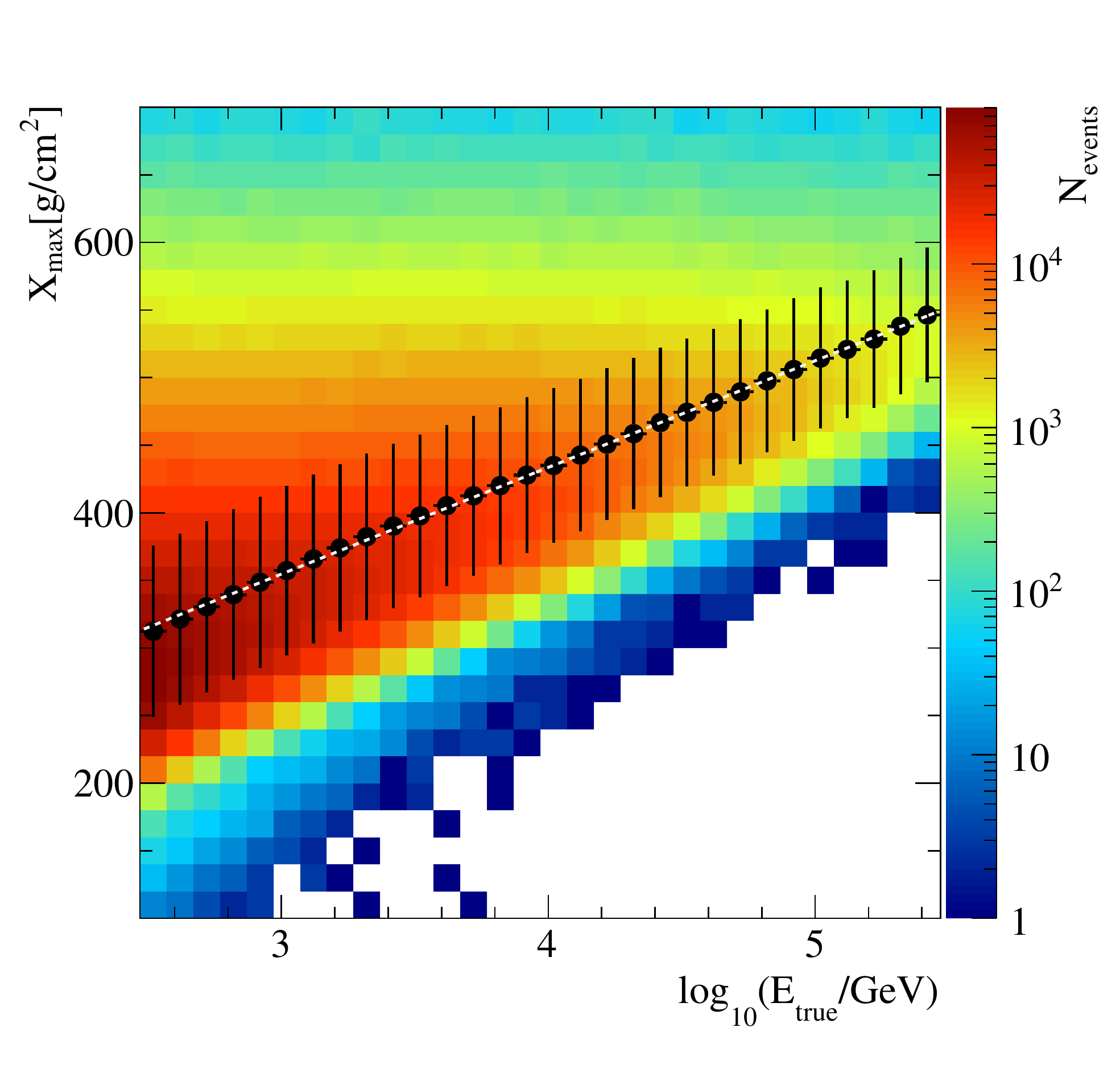}
\caption{The left panel shows the relation between the number of tanks got at least a single pe hit and the distance of the shower core from the centre of the array. The true energy (E$_{\rm true}$) of the primary $\gamma$-ray photon is shown on the z-axis. The right panel is showing the relation between the E$_{\rm true}$ and the true X$_{\rm max}$. The white dashed line shows the functional relation. The z-axis shows the number of shower events (N$_{\rm events}$).}
\label{fig:energy and xmax guess}
\end{figure*}

To start the minimisation with a reasonable value of $E$ we use the information of the number of detectors that got at least a single pe hit (N$_{\rm hit}$) as a function of the distance of the core location from the centre of the array.  This distribution we show in Figure \ref{fig:energy and xmax guess} (left panel), one can estimate the energy of a given shower by using the value of the observables N$_{\rm hit}$ and the distance of the trial core from the centre of the array. $E_{\rm true}$ denotes the simulated energy of the primary $\gamma$-ray photon.

As explained in Section \ref{considerations}, one of the challenges is finding the  global minimum while there are ambiguities in LDF for different values of $X_{\text{max}}$ and shower energy. Therefore, in the first pass of the fit procedure, we enforce the linear relation between $E_{\text{true}}$ and  the average value of $X_{\rm max}$ as shown in Figure \ref{fig:energy and xmax guess} (right panel).

To reduce the total computation time, the number of iterations of the minimiser is limited to 10 for each of the three starting points. From  the three resulting minimum likelihood values, we select the smallest value and the corresponding fit parameters are used as the starting point in the second pass of the fit method.

\subsection{Second Pass}
The results of the first pass of the fit-procedure are passed to the direction-reconstruction method, which can get a more accurate estimate of the direction now that it has a better knowledge of the shower core location.  The results from the direction-reconstructor are used, together with the location of the shower core as the starting point for the next pass. In this pass, the constrained on the relation between $X_{\text{max}}$ and $E_{\text{true}}$ is no longer enforced and the maximum number of iterations of the minimiser is increased to 40. Typically the fit procedure converges on the best fitting parameters before reaching the maximum number of iterations. The reconstructed core now can further be used to improve the direction-reconstruction.

\subsection{Examples}
\begin{figure*}[!h]
\centering
\includegraphics[width=\linewidth]{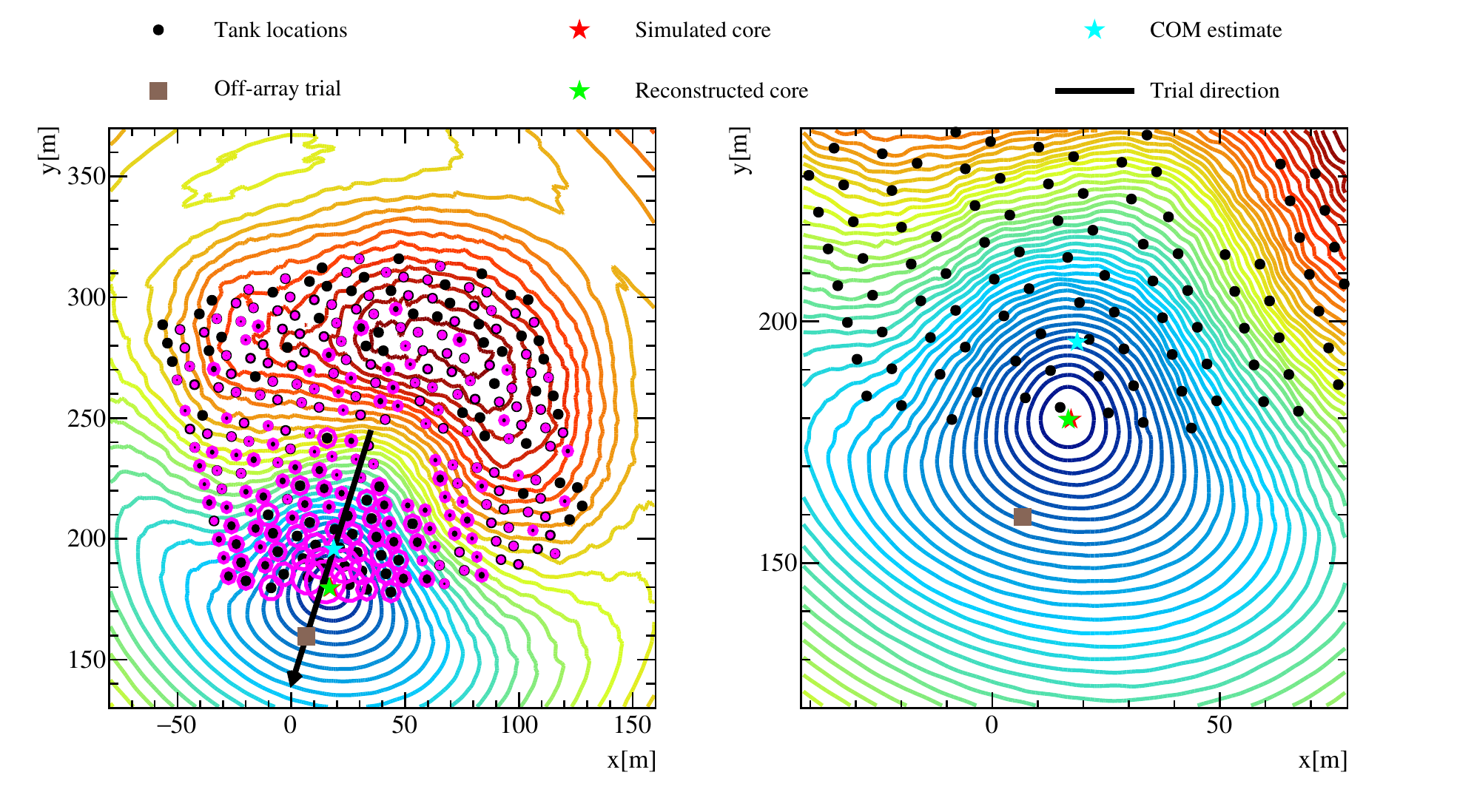}
\caption{Example MC event. The figure on the right is the zoom in version of the figure on the left around the reconstructed core. The likelihood surface contours from blue to red colour show the minimum to maximum respectively. The magenta circles over the tanks show the relative charge observed between the different tanks.}
\label{fig:likelihood_surface_HAWC300}
\end{figure*}

We show one typical example MC event of the $\gamma$-ray reconstructed energy of $\sim$ 17 TeV in Figure \ref{fig:likelihood_surface_HAWC300}. 
The 2D projection of the likelihood surface shows the contours with different colours. The contour colour scale varies from red to blue indicating the maximum to minimum of the likelihood surface respectively. The red colour star shows the true (simulated) core location. The COM location is shown with the cyan colour star.  The magenta colour circles over the tanks denote the relative charge observed between the tanks. We also show the off-array trial direction and one such example. The green colour star shows the reconstructed core as the best fit result which falls in the prominent minimum shown with blue coloured contours of the likelihood surface. It can be seen that the method converged to very close to the true core location. 

\begin{figure*}[!h]
\centering
\includegraphics[width=0.49\linewidth]{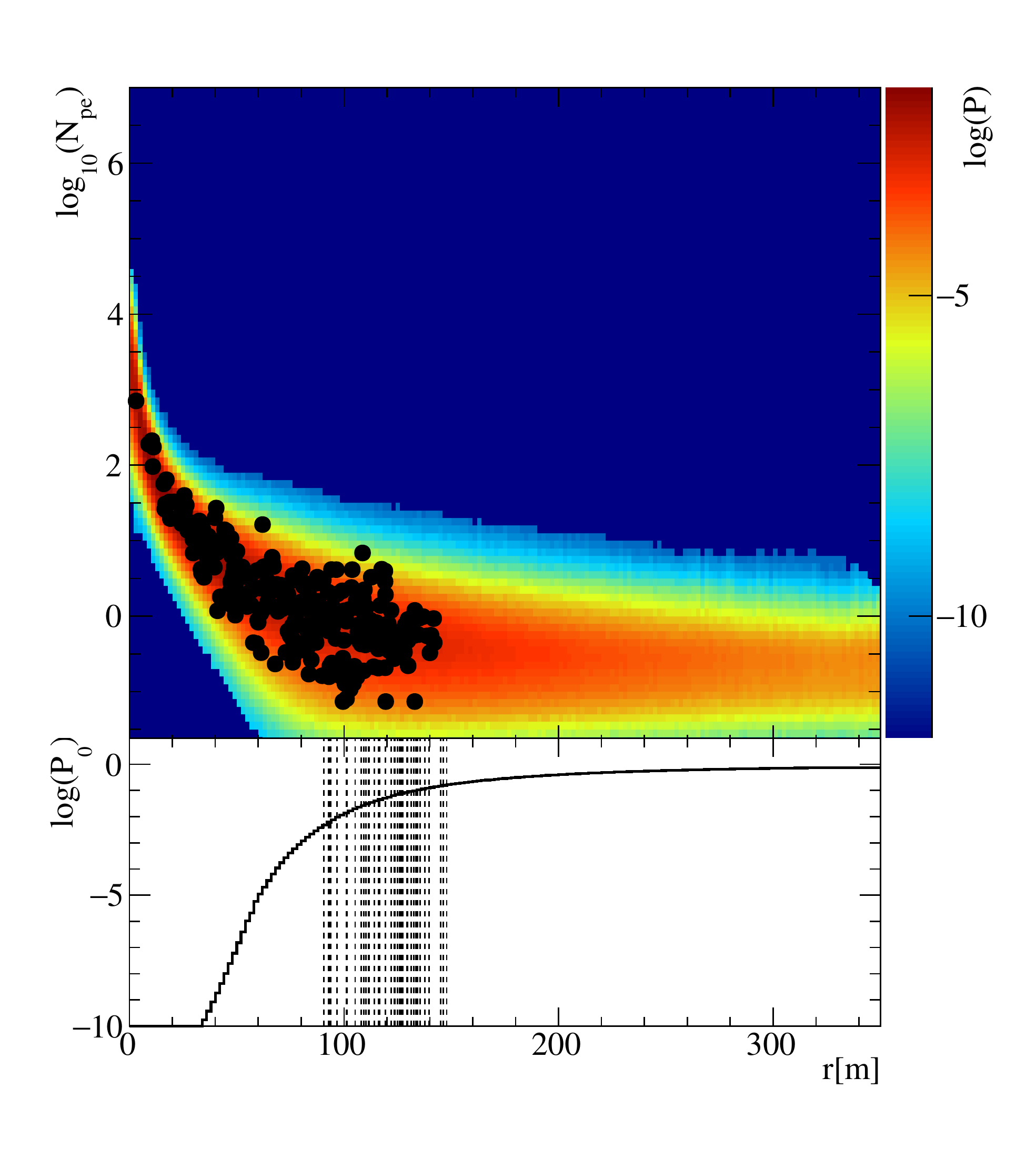}
\includegraphics[width=0.49\linewidth]{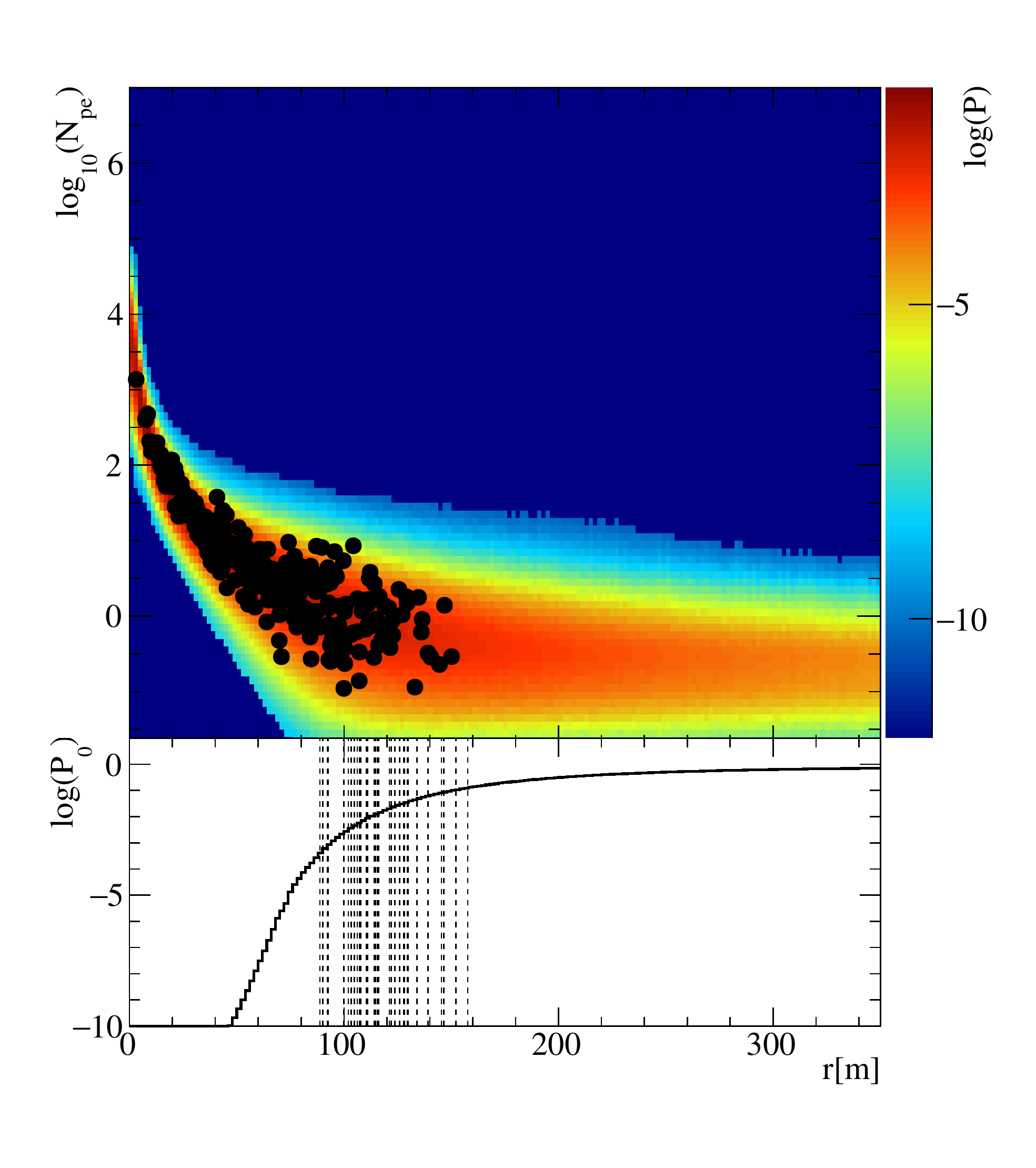}
\caption{The LDF and PDF template corresponding to the event in Figure \ref{fig:likelihood_surface_HAWC300} with true energy $\sim$ 16 TeV, reconstructed energy of $\sim$ 17 TeV and true X$_{\rm max}$ $\sim$ 400 g/cm$^{2}$, reconstructed X$_{\rm max}$ $\sim$ 423 g/cm$^{2}$ and zenith angle $\sim$ 20.76$^{\circ}$ for the MC event (left panel). One similar Crab Nebula data event  is also shown (right panel) with the reconstructed energy of $\sim$ 17 TeV, X$_{\rm max}$ $\sim$ 424 g/cm$^{2}$ and zenith angle $\sim$ 4.03$^{\circ}$. The black dots show the LDFs for non-zero N$_{\rm pe}$ and observed zeros are shown as the dashed lines on the histogram below with their corresponding probability. }
\label{fig:LDF_PDF_HAWC300}
\end{figure*}

In Figure \ref{fig:LDF_PDF_HAWC300} (left panel) we show the LDF of the example MC $\gamma$-ray shower discussed above and the template PDF corresponding to the best fit parameter values. The black dots represent the tank locations and observed signal and the dashed line show the observed zero pe. Similarly, as an example applied to real data, we show in Figure \ref{fig:LDF_PDF_HAWC300} (right panel) the best fitting template PDF and LDF of a data event measured by HAWC, which is selected after hadron rejections cuts from the direction of the Crab Nebula.  

\section{Binning Optimisation}
\label{Optimisation of Binning}
Deciding the bin size for different fit parameters to make the templates is a crucial step. On one hand, too fine binning can be very computing intensive for making the templates and performing  the fit, while on the other hand, too coarse binning affects the smoothness of the likelihood function and  hence will decrease the achievable resolutions of the fitted parameters.

To find an optimal bin size for each parameter, we created templates of varying bin sizes for a given parameter while fixing the others.  Typically, reducing the bin size in a parameter will improve the resolution and reduce the bias up to a point where binning is not the dominant effect on parameter estimation anymore. 

Firstly we fix the shower energy at 10\,TeV, and we optimise the binning in $X_{\text{max}}$ and $r$. We have chosen 10\,TeV as being rather central value in the sensitivity of air shower detector arrays like HAWC, however, we note here that the procedure of optimisation can be repeated at different energies. 

\begin{figure}[!h]
\centering
\includegraphics[width=0.7\linewidth]{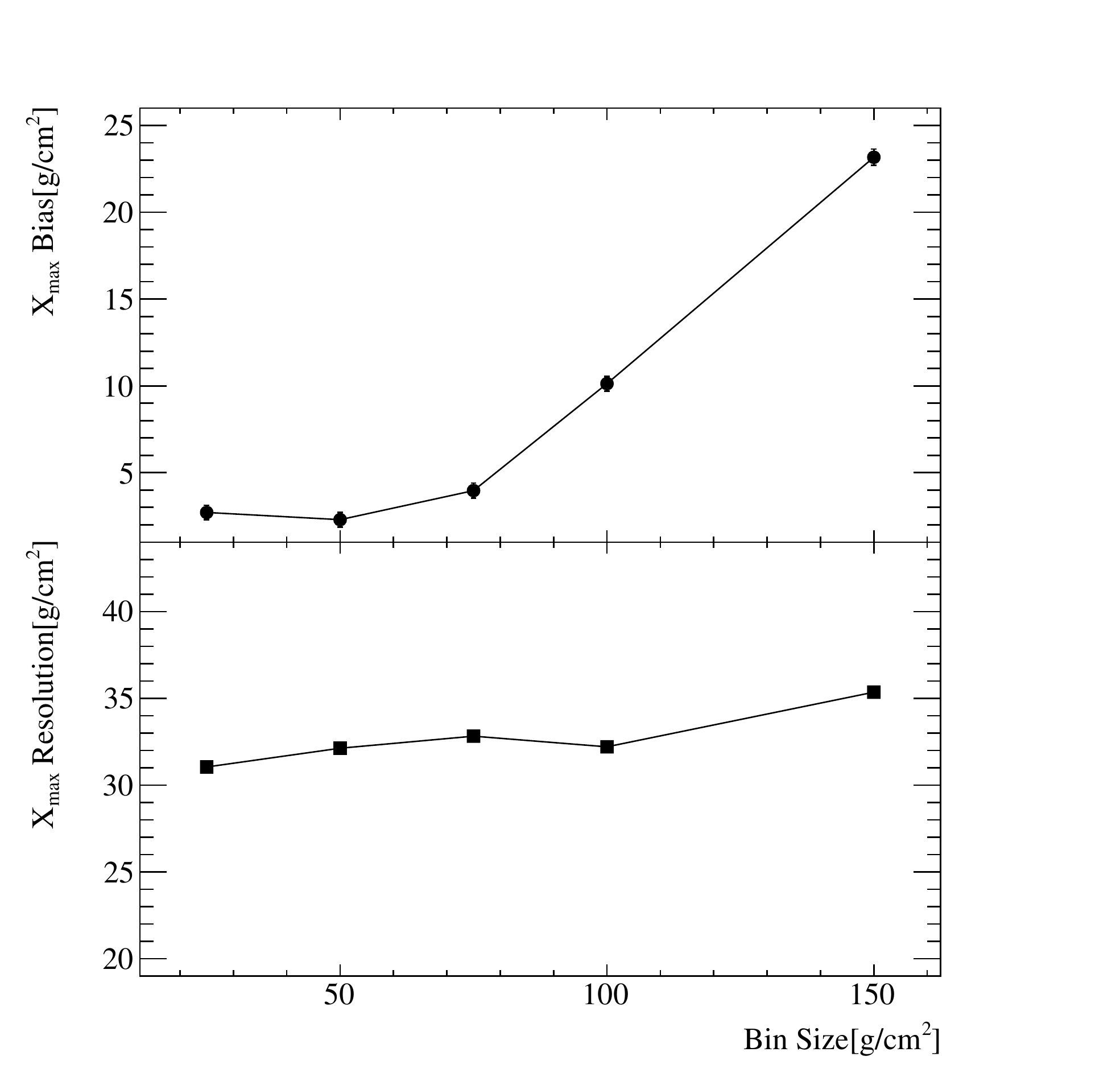}
\caption{X$_{\rm max}$ bias ($\langle X_{\text{max,reco}} -X_{\text{max,true}} \rangle$) and X$_{\rm max}$ resolution  ($\text{RMS}(X_{\text{max,reco}} -X_{\text{max,true}})$) shown as a function of bin size of X$_{\rm max}$.}
\label{fig:Xmax_binning}
\end{figure}
In Figure  \ref{fig:Xmax_binning} we show the case of where we varied bin sizes (25, 50, 75, 100 and 150\,g/cm$^2$) of  $X_{\rm max}$ and fixed the bin size of energy to 0.1 in $\log_{10}$(E/GeV) and  core distance to 2 m. While fitting $X_{\rm max}$, we fixed the other two parameters to their simulated values and only let $X_{\rm max}$ as a free parameter. As $X_{\rm max}$ is also dependent on energy, we have chosen only 10 TeV energy showers. Figure \ref{fig:Xmax_binning}, shows the $X_{\rm max}$ bias and  $X_{\rm max}$ resolution as a function of $X_{\rm max}$  bin size. $X_{\rm max}$ bias and $X_{\rm max}$ resolution are the mean and RMS of  $X_{\rm max,reco}$ - $X_{\rm max,true}$ distribution, which is approximately a Gaussian. Reducing the bin size below 50 g/cm$^2$ seems not to improve resolution and bias any further, therefore we chose it as the optimal value. 

\begin{figure}[!h]
\centering
\includegraphics[width=0.7\linewidth]{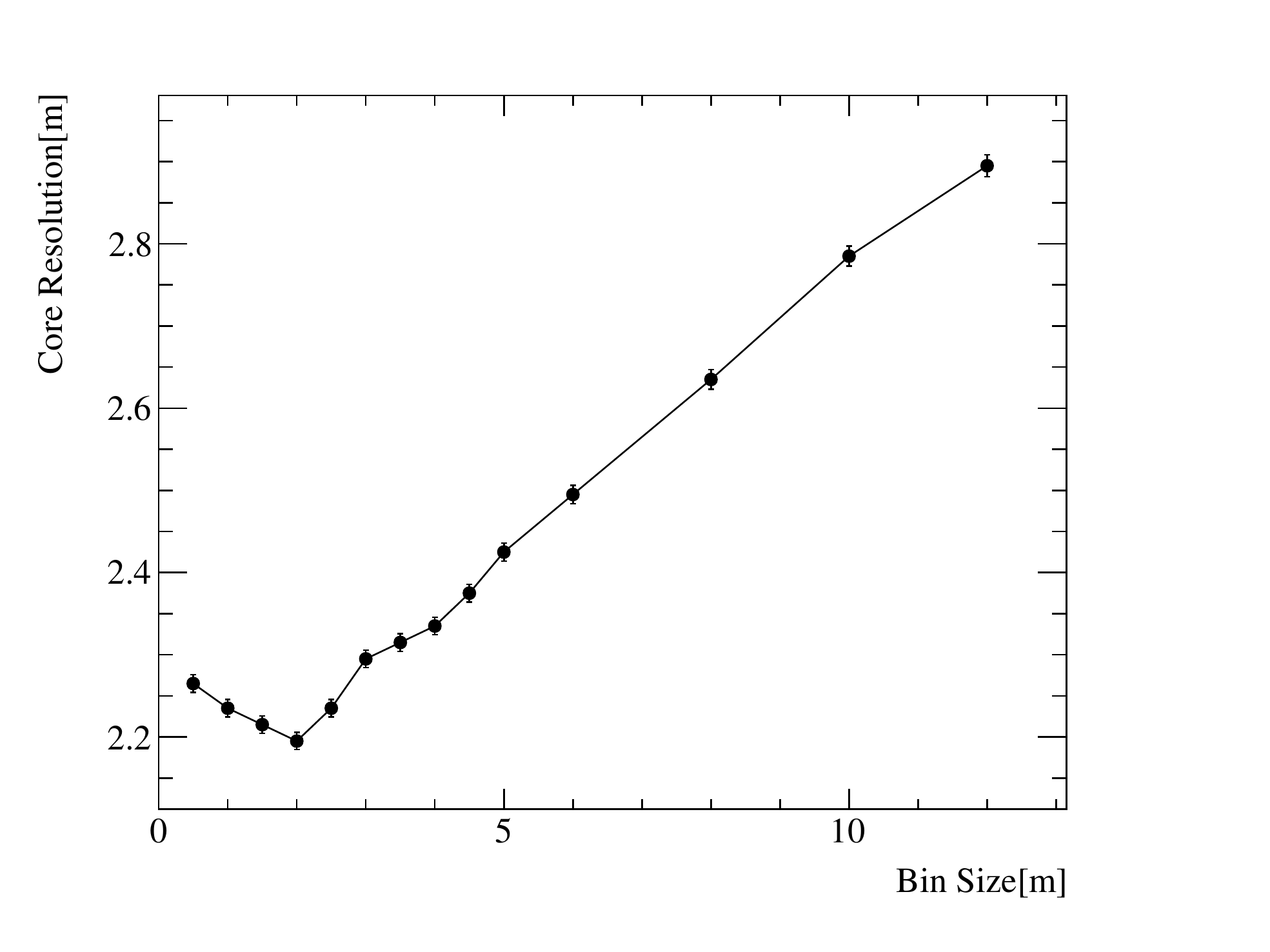}
\caption{Core resolution (68\% containment radius) versus varying bin size of distance from the shower core.}
\label{fig:core_binning}
\end{figure}

We repeated the procedure for the $r$ binning. Figure \ref{fig:core_binning} shows that the optimal bin size at 10 TeV energy was 2 m. The points for a given bin size represents the core resolution (68\% containment radius) of the corresponding core (true) - core (reco) distribution. 
\begin{figure}[!h]
\centering
\includegraphics[width=0.7\linewidth]{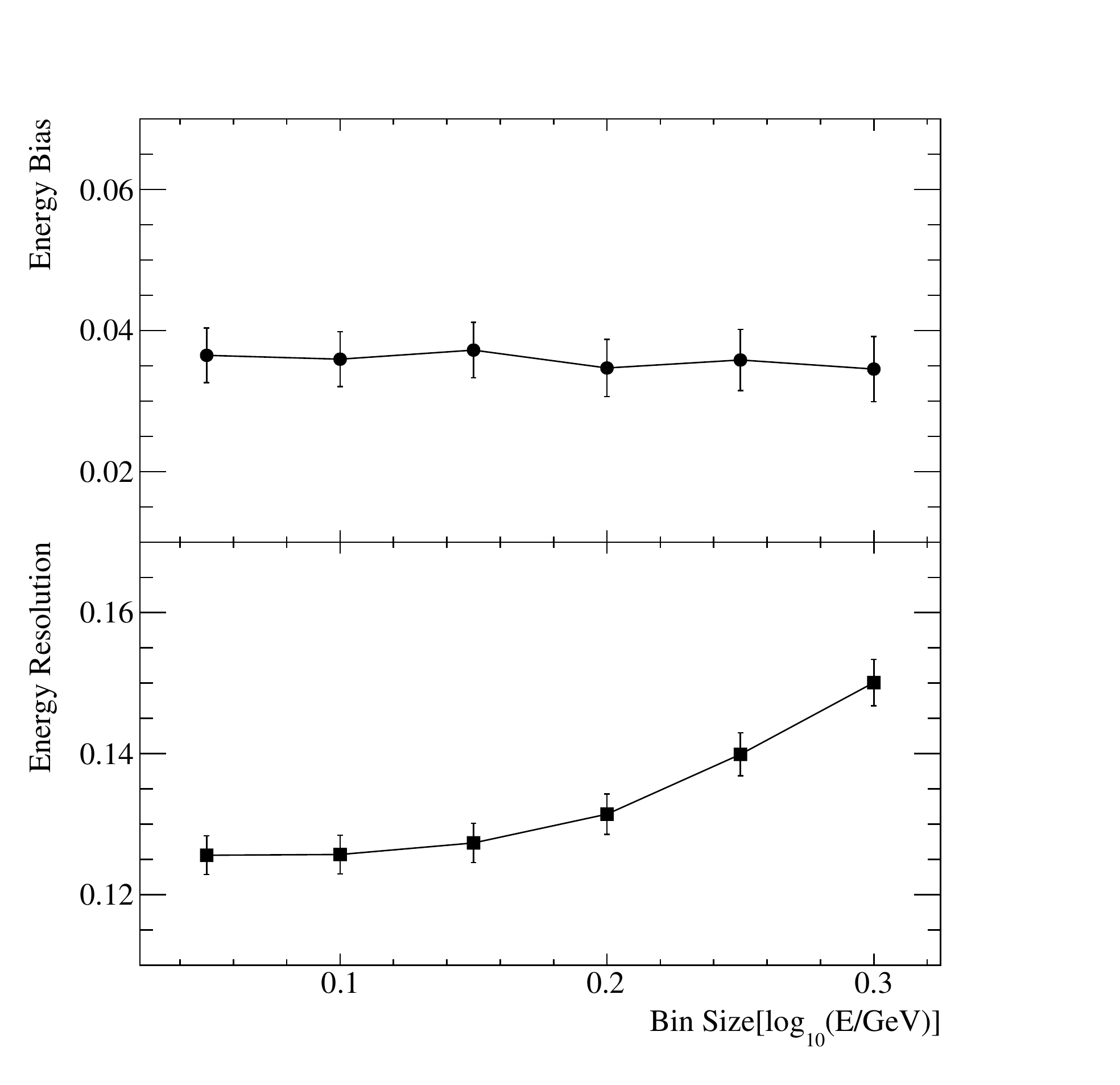}
\caption{Fractional energy bias (top) and energy resolution (bottom) versus bin size of true energy of the $\gamma$-ray photon.}
\label{fig:Energy_binning}
\end{figure}

To optimise the $E$ bin size, we again repeated the procedure for creating the templates, but for the test sample, we took an energy range instead of only 10 TeV energy to escape the discreteness in the $\log_{10}$(E$_{\rm reco}$) - $\log_{10}$(E$_{\rm true}$) distribution. From the fit output for the energy range, we took the 10 TeV bin, which includes some nearby energies and makes the distributions smooth. By looking at Figure \ref{fig:Energy_binning}, 0.1 in $\log_{10}$(E/GeV) seems to be the best choice.

\section{Performance}
\label{perforamance}
We assessed the performance of the fit method using MC simulations for the HAWC array. The test simulation data set has the same ranges for the parameters $E$, $X_{\rm max}$ and $\theta$ as of the simulation dataset for generating the MC templates which are described in Section \ref{simulation dataset}. To have events of a reasonable size, we put an additional condition that at least 10\% of the available channels should have observed a signal. 

We show the performance on the core and energy reconstruction and gamma-hadron separation. The performance on X$_{\rm max}$ is not shown because it was found to be dominated by the prior relation we have used between $X_{\rm max}$ and $E$ as described in Section \ref{likelihood fitting}. However, it is to be noted that the template binning and fitting in X$_{\rm max}$ is crucial in order to partially break the ambiguity in X$_{\rm max}$ and energy as discussed in Section \ref{ambiguity in Energy and Xmax}.

\subsection{Core Resolution}

\label{core resolution}
\begin{figure}[!h]
\centering
\includegraphics[width=0.7\linewidth]{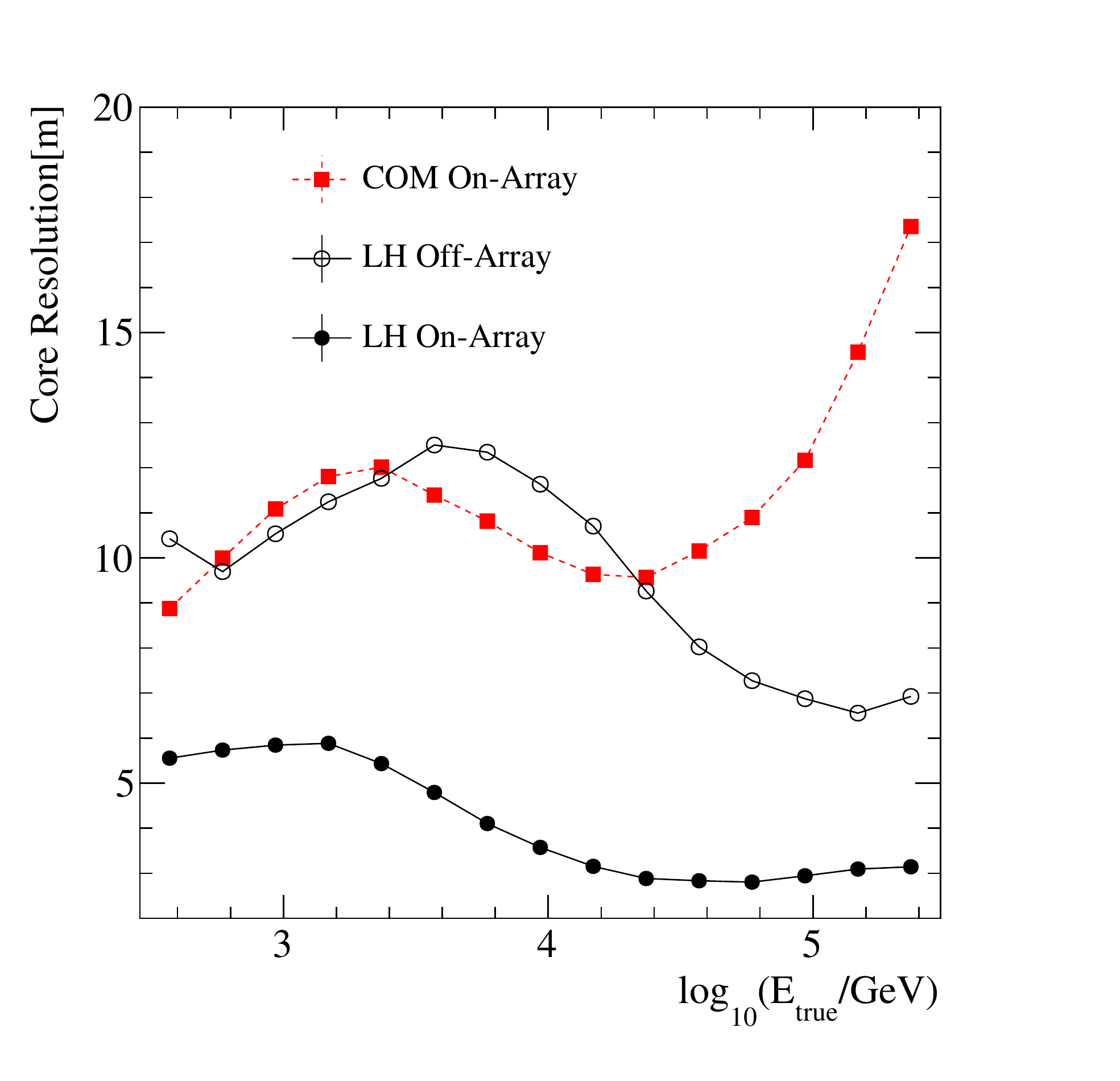}
\caption{Core resolution (68\% containment radius) shown as a function of the true energy of the $\gamma$-ray. Here COM and LH are centre of mass estimate and likelihood fit respectively. Off and On-Array stands for the true core of the shower falling inside and outside the array respectively.}
\label{fig:result_core_reso}
\end{figure}
In figure \ref{fig:result_core_reso}, the core resolution as a function of true $\gamma$-ray energy is shown. We defined the core resolution as 68\% containment radius of the distribution of the distance between the reconstructed and true shower core. To give a reference, we show the core resolution for a simple Centre-Of-Mass (COM) estimation (red markers). We show the core resolution for the true core of the showers falling inside (solid markers) and outside (hollow markers) the array. It can be seen over all the energy range the likelihood (LH) procedure works better than the simple COM estimation as expected. The core resolution is $\sim$6 m for 0.3 TeV and then falls down to $\sim$3 m for energies larger than 10 TeV for the events falling inside the array. For events falling outside the array over a radially symmetric area, which is 50\% as of the area of the main array the core resolution is $\sim$10 m at 0.3 TeV and again falls down to $\sim$7 m for the highest energies. It is worth to note that even for the events falling outside the array the LH performs similar at the low energies and better at high energies in comparison to the COM on-array estimate.

Overall, the shower core is accurately reconstructed over the whole energy range. At the low energies, even though there are fewer particles observed on the ground, we get reasonable fit results which can be partially attributed to the fact that method takes into account the zero pe signal. 

\subsection{Energy Resolution and Bias}
\label{energy resolution and bias}
\begin{figure}[!h]
\centering
\includegraphics[width=0.7\linewidth]{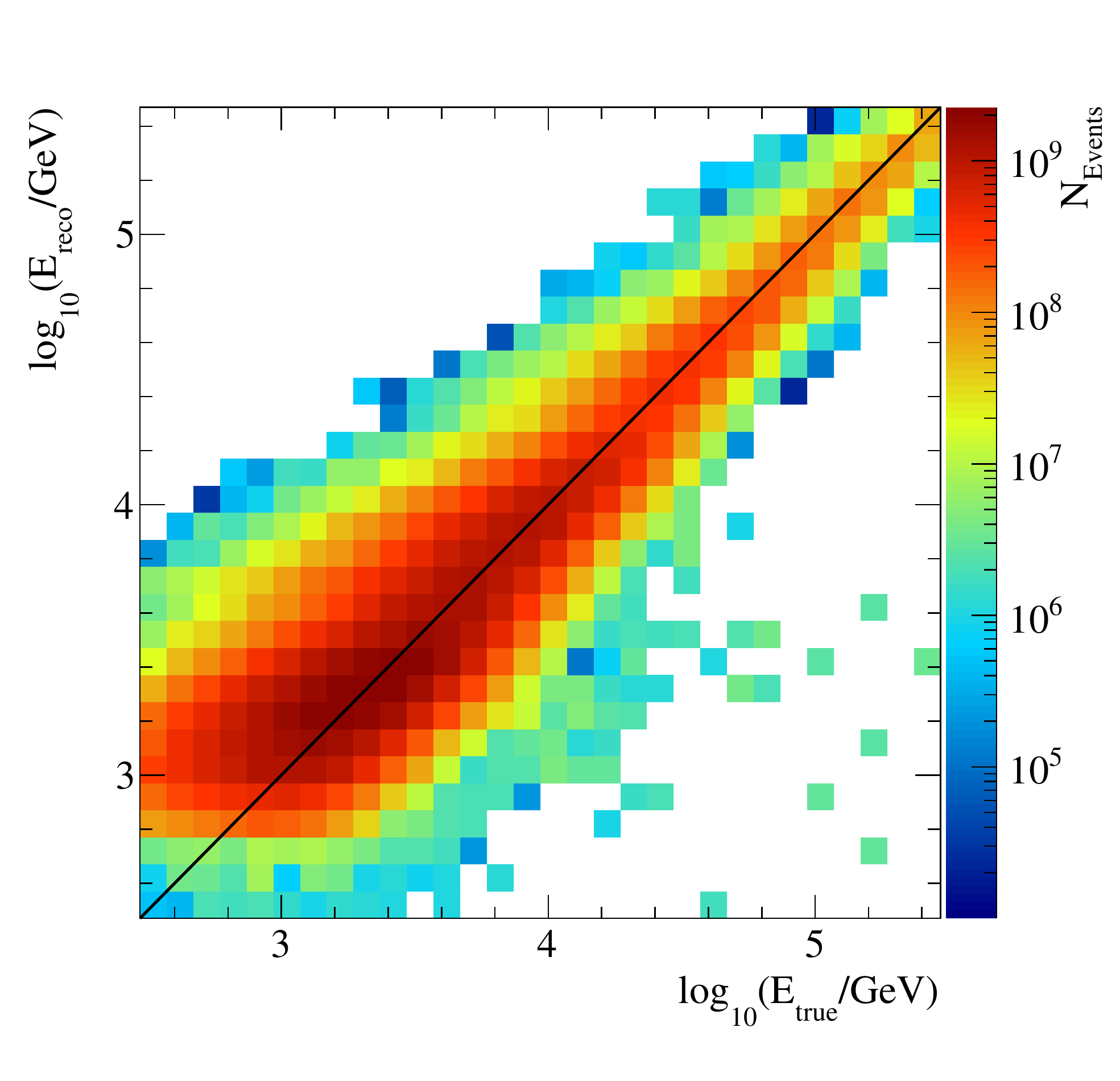}
\caption{Distribution of reconstructed $\gamma$-ray photon energy versus true $\gamma$-ray photon energy. The z-axis shows the number of $\gamma$-ray events.}
\label{fig:energy_true_vs_reco}
\end{figure}

\begin{figure}[!h]
\centering
\includegraphics[width=0.7\linewidth]{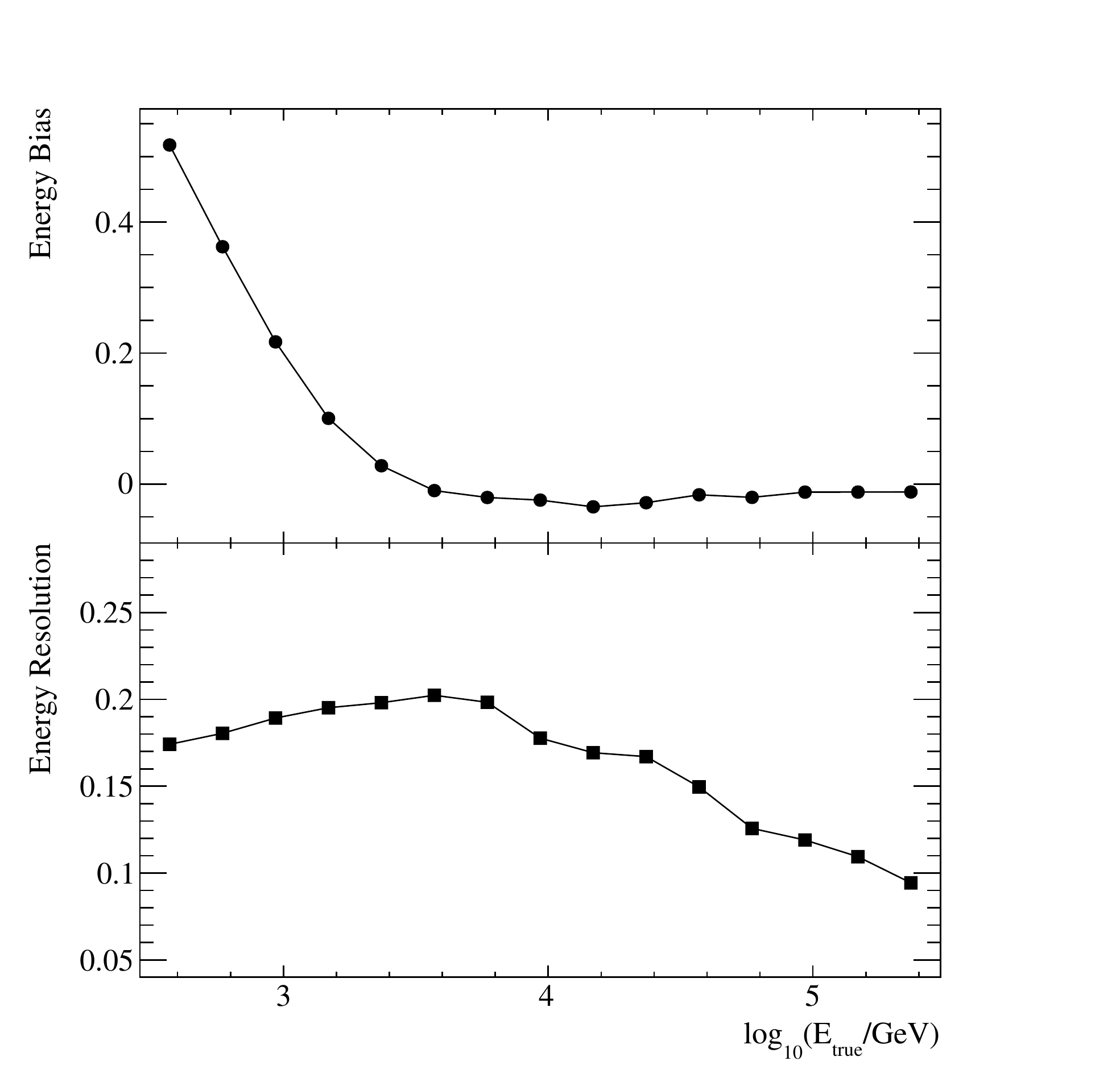}
\caption{Fractional energy bias (top) and energy resolution (bottom) as a function of true $\gamma$-ray photon energy.}
\label{fig:energy_mean_rms}
\end{figure}

In Figure \ref{fig:energy_true_vs_reco} we show the distribution of reconstructed energy versus true energy. The z-axis shows the number of shower events. The figure shows the strong correlation between reconstructed and true energy.

The performance of the energy reconstruction can be evaluated by with the fractional deviation ($\log_{10}$(E$_{\rm reco}$) - $\log_{10}$(E$_{\rm true}$)) of the reconstructed energy with respect to the true energy. The deviation from 0 of the mean of the distribution of this fractional deviation (see Figure \ref{fig:energy_mean_rms} (top panel)) is the bias in the energy reconstruction. Similarly, the RMS can be understood as the energy resolution(see Figure \ref{fig:energy_mean_rms} (bottom panel)). 

The energy bias is large at the low energies but converges to zero at energies $>$4\,TeV, which indicates the stable region of the energy reconstruction will start above this value. The bottom panel of Figure \ref{fig:energy_mean_rms} shows the energy resolution of the likelihood fit method, which starts at $\sim$50\% for low energy and drops down to $\sim$25\% at the highest energies. Such energy resolution is very promising for EAS arrays like HAWC.

\subsection{Gamma-Hadron Separation}

\begin{figure}[!h]
\centering
\includegraphics[width=0.7\linewidth]{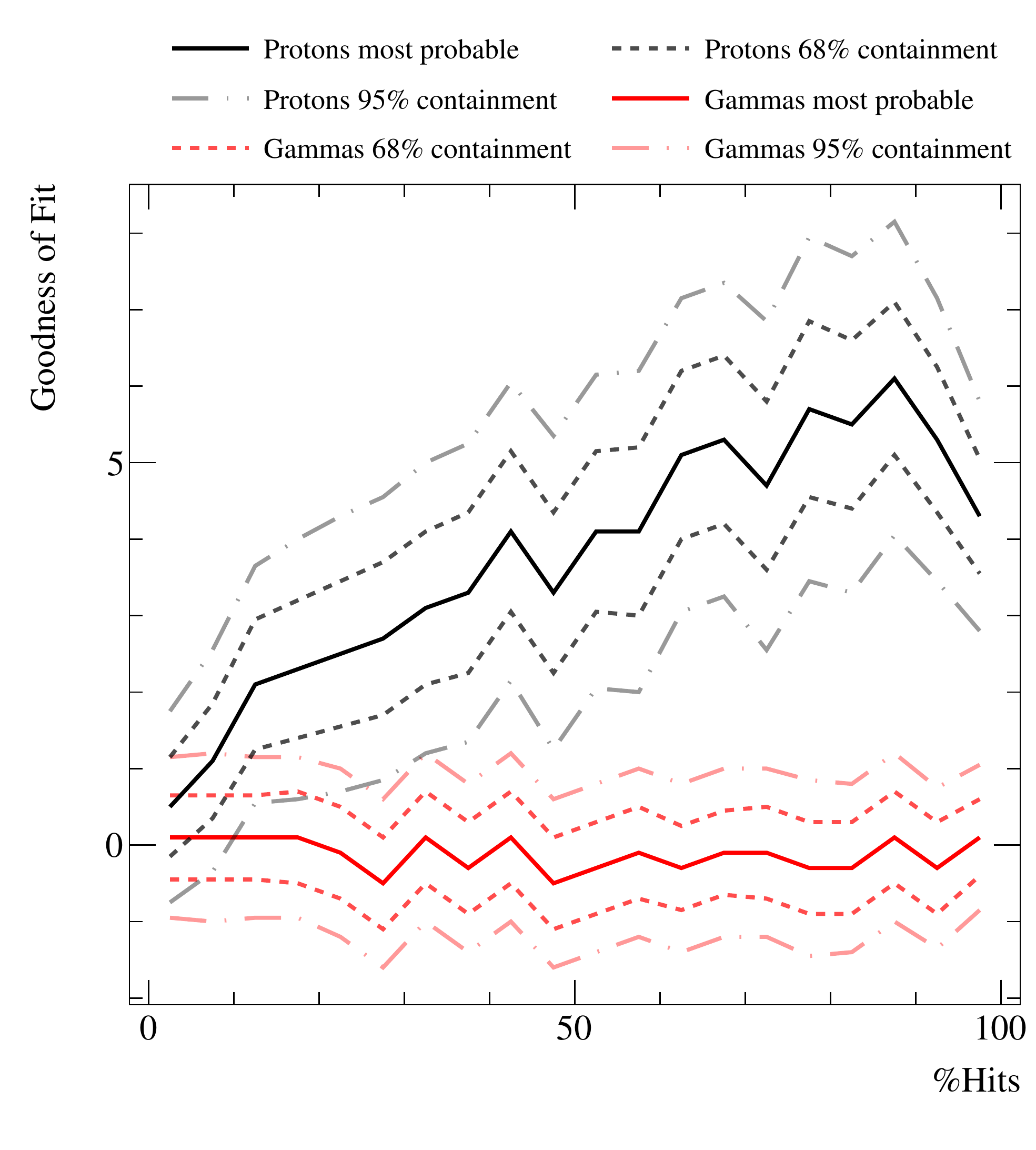}
\caption{The GoF distributions of the likelihood fit method for simulated $\gamma$-ray and proton induced air shower are shown in red and black respectively. It shows the most probable value of GoF with 68\% and 95\% containment as a function of percentage hits of the HAWC main array.}
\label{fig:gof_vs_percentage_hits}
\end{figure}

The concept of the goodness of fit is explained in Section \ref{likelihood function}. To demonstrate the gamma-hadron separation power, we show the GoF distributions for the simulated MC air showers. 
Figure \ref{fig:gof_vs_percentage_hits} exhibits the GoF distributions as a function of observed percentage hits (percent number of detector units) of the HAWC array. It shows the most probable value of the GoF distribution with 68\% and 95\% containment. We present, the distributions for simulations of $\gamma$-ray and proton induced air showers. For the higher mass hadrons, one would expect even more diverge distribution of the GoF from the $\gamma$-ray induced air showers. 

It is evident from the GoF distribution of $\gamma$-ray and proton induced air showers that it starts showing a separation power for as low as $\sim$20\% of the percentage hits of the array and the separation power increase with event size. The last part in the graph shows a slight decrease in separation power, which might be due to saturation effects of the detector.

\section{Application to Mixed Detector Arrays}
\label{application to mixed type particle detector arrays}

\begin{figure*}[!h]
\centering
\includegraphics[width=\linewidth]{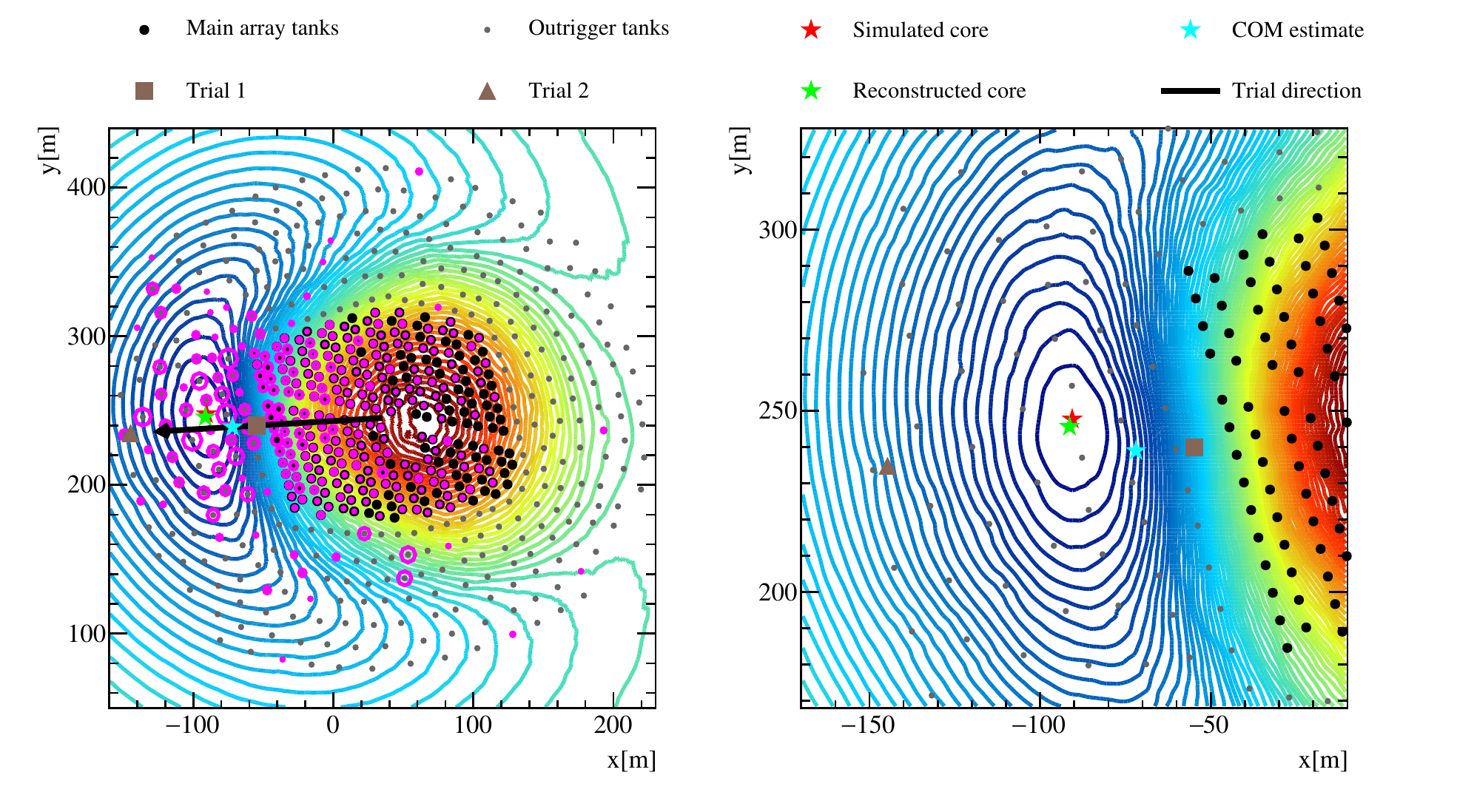}
\caption{The figure on the right is the zoom in version of the figure on the left around the reconstructed core. It shows the likelihood surface contours for an example event. Blue to red colour contours show the minimum and the maximum of the surface respectively. The magenta circles over the tanks show the relative charge observed between the different tanks.}
\label{fig:likelihood_surface_HAWC+OR}
\end{figure*}

In this section, we illustrate the applicability of this method for mixed detector type air shower arrays by using detector simulations for HAWC and its high-energy upgrade, the outrigger array \cite{hawc_Outriggers1,hawc_Outriggers2}. The outrigger array consists of 345 smaller Water Cherenkov Detectors (WCDs) around the main array of the HAWC $\gamma$-ray observatory (see Figure \ref{fig:likelihood_surface_HAWC+OR} for the schematic layout). The typical separation between the two outrigger WCDs is $\sim$15 m. One outrigger WCD is consists of a cylindrical tank of 1.55 m diameter and 1.65 m height equipped with one 8" PMT at the bottom of the tank, and therefore its response is significantly different than the WCDs of the HAWC main array. As explained in Section \ref{likelihood function}, it is straightforward to combine different detector types in the likelihood function.  

Figure \ref{fig:likelihood_surface_HAWC+OR} an example is illustrated of the  method applied to a simulated event with signals detected in both WCDs of outriggers and the main array of HAWC.
\begin{figure*}[!h]
\centering
\includegraphics[width=0.49\linewidth]{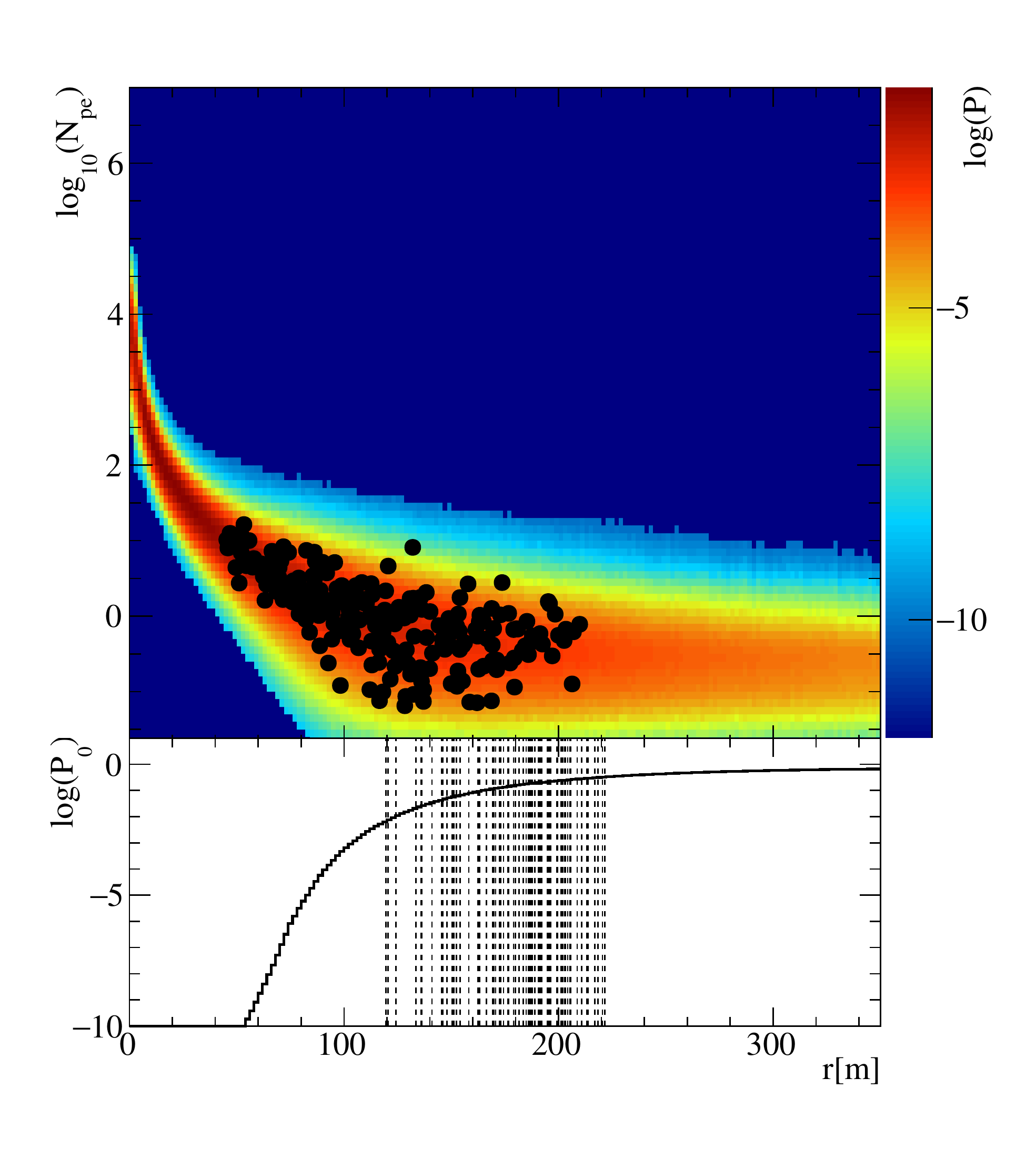}
\includegraphics[width=0.49\linewidth]{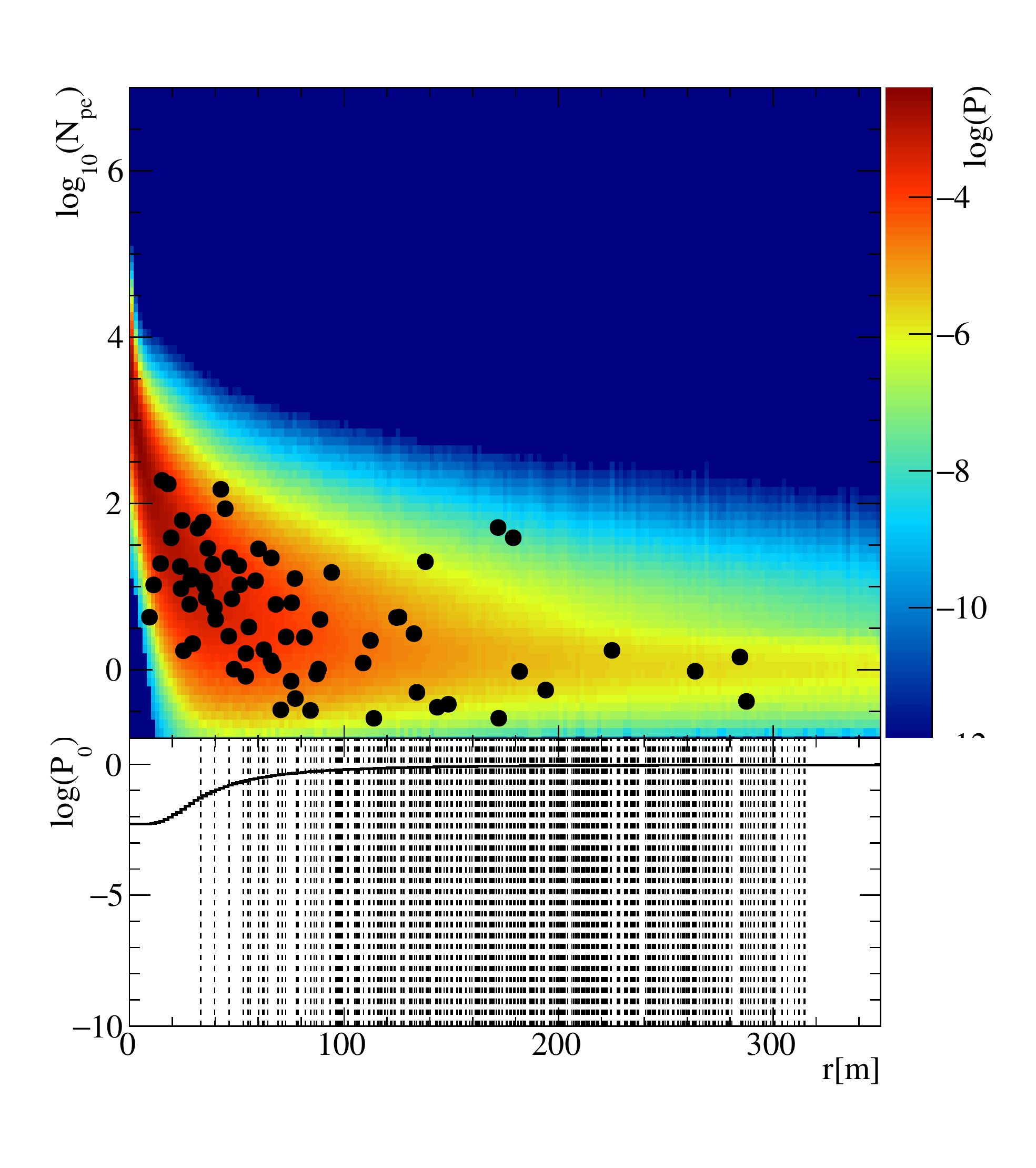}
\caption{The PDF templates (main array tanks: left, outrigger tanks: right) corresponding to the fit values of $E$, $X_{\rm max}$ and $\theta$ for a typical event shown in  Figure \ref{fig:likelihood_surface_HAWC+OR}. The black dots show the LDF for non-zero N$_{\rm pe}$ and observed zeros are shown as the dashed lines on the histograms below with their corresponding probability.}
\label{fig:LDF_PDF_HAWC+OR}
\end{figure*} 
It shows the likelihood contours and the reconstructed and the true core for a simulated event located in the area instrument by the outrigger array. The  best fitting templates corresponding to the main array tanks and the outriggers are shown in Figure \ref{fig:LDF_PDF_HAWC+OR} left and right panel respectively. In the case of the smaller outrigger detectors, the fluctuations in the particle density of the shower front become more prominent which manifest in both larger amplitude fluctuations and more zero signal detectors. Nevertheless, it is not a problem for the template-based likelihood fit procedure, which combines these significantly different detector responses naturally to reconstruct the $\gamma$-ray properties.

\section{Conclusion and Prospects}
\label{conclusions and future prospects}
We described the working of the MC template-based likelihood fit method and showed the possibility of $\gamma$-ray shower reconstruction for air shower arrays. This work demonstrates the reconstruction of core location and the energy of the primary particle for $\gamma$-ray induced air showers using the observed LDF. The goodness of fit of the method  shows that in addition to $\gamma$-ray reconstruction, it can be used to reject the background from hadron induced air showers.  In principle, this method can be expanded for direction reconstruction while using the timing of observed signals. Although we have shown its working only for $\gamma$-ray induced air showers, it might be possible to apply it to the reconstruction of other particle species. However, with the mixed composition of hadronic induced cosmic-ray air showers and the uncertainty on hadronic interaction models, it might be more challenging to obtain reconstruction with similar accuracy.

Currently, this method is in consideration for reconstruction of the HAWC data. It can both be applied to archival data and to the new combined dataset to be taken with both the main array and the high-energy upgrade with the outrigger array.

\section*{Acknowledgements}
\label{acknowledgements}
We thank the HAWC collaboration for granting access to the HAWC detector simulation chain and data sample and for fruitful discussions. We would also like to thank Dr. Vincent Marandon for useful discussions and help.  We gratefully acknowledge financial support from the International Max Planck Research School for Astronomy and Cosmic Physics at the University of Heidelberg and the Max Planck Society, Germany. 

%\section*{References}
%\label{references}
\bibliographystyle{unsrt}
\bibliography{bib_file}

\begin{thebibliography}{10}

\bibitem{ref_NKG}
K.~{Kamata} and J.~{Nishimura}.
\newblock {The Lateral and the Angular Structure Functions of Electron
  Showers}.
\newblock {\em Progress of Theoretical Physics Supplement}, 6:93--155, 1958.

\bibitem{ref_cosmic_ray_showers}
K.~{Greisen}.
\newblock {Cosmic Ray Showers}.
\newblock {\em Annual Review of Nuclear and Particle Science}, 10:63--108,
  1960.

\bibitem{ref_CAT1}
A.~Barrau et~al.
\newblock {The CAT imaging telescope for very high-energy gamma-ray astronomy}.
\newblock {\em Nucl. Instrum. Meth.}, A416:278--292, 1998.

\bibitem{ref_CAT2}
S.~Le~Bohec et~al.
\newblock {A New analysis method for very high definition imaging atmospheric
  Cherenkov telescopes as applied to the CAT telescope}.
\newblock {\em Nucl. Instrum. Meth.}, A416:425--437, 1998.

\bibitem{ref_CAT_method_in_HESS}
M.~{de Naurois} and L.~{Rolland}.
\newblock {A high performance likelihood reconstruction of {$\gamma$}-rays for
  imaging atmospheric Cherenkov telescopes}.
\newblock {\em Astroparticle Physics}, 32:231--252, December 2009.

\bibitem{ref_IMPACT}
R.~D. {Parsons} and J.~A. {Hinton}.
\newblock {A Monte Carlo template based analysis for air-Cherenkov arrays}.
\newblock {\em Astroparticle Physics}, 56:26--34, April 2014.

\bibitem{ref_HESS}
J.~A. Hinton.
\newblock {The Status of the H.E.S.S. project}.
\newblock {\em New Astron. Rev.}, 48:331--337, 2004.

\bibitem{hawc_all_particle_spectrum}
R.~Alfaro et~al.
\newblock {All-particle cosmic ray energy spectrum measured by the HAWC
  experiment from 10 to 500 TeV}.
\newblock {\em Phys. Rev.}, D96(12):122001, 2017.

\bibitem{ref_hawc}
A.~U. Abeysekara et~al.
\newblock {Sensitivity of the High Altitude Water Cherenkov Detector to Sources
  of Multi-TeV Gamma Rays}.
\newblock {\em Astropart. Phys.}, 50-52:26--32, 2013.

\bibitem{hawc_Outriggers2}
Vikas Joshi and Armelle Jardin-Blicq.
\newblock {HAWC High Energy Upgrade with a Sparse Outrigger Array}.
\newblock In {\em {Proceedings, 35th International Cosmic Ray Conference (ICRC
  2017): Bexco, Busan, Korea, July 12-20, 2017}}, 2017.

\bibitem{Tibet_array}
M.~Amenomori, X.~J. Bi, D.~Chen, T.~L. Chen, W.~Y. Chen, S.~W. Cui,
  Danzengluobu, L.~K. Ding, C.~F. Feng, Zhaoyang Feng, Z.~Y. Feng, Q.~B. Gou,
  Y.~Q. Guo, H.~H. He, Z.~T. He, K.~Hibino, N.~Hotta, Haibing Hu, H.~B. Hu,
  J.~Huang, H.~Y. Jia, L.~Jiang, F.~Kajino, K.~Kasahara, Y.~Katayose, C.~Kato,
  K.~Kawata, M.~Kozai, Labaciren, G.~M. Le, A.~F. Li, H.~J. Li, W.~J. Li,
  C.~Liu, J.~S. Liu, M.~Y. Liu, H.~Lu, X.~R. Meng, T.~Miyazaki, K.~Mizutani,
  K.~Munakata, T.~Nakajima, Y.~Nakamura, H.~Nanjo, M.~Nishizawa, T.~Niwa,
  M.~Ohnishi, I.~Ohta, S.~Ozawa, X.~L. Qian, X.~B. Qu, T.~Saito, T.~Y. Saito,
  M.~Sakata, T.~K. Sako, J.~Shao, M.~Shibata, A.~Shiomi, T.~Shirai,
  H.~Sugimoto, M.~Takita, Y.~H. Tan, N.~Tateyama, S.~Torii, H.~Tsuchiya,
  S.~Udo, H.~Wang, H.~R. Wu, L.~Xue, Y.~Yamamoto, K.~Yamauchi, Z.~Yang,
  S.~Yasue, A.~F. Yuan, T.~Yuda, L.~M. Zhai, H.~M. Zhang, J.~L. Zhang, X.~Y.
  Zhang, Y.~Zhang, Yi~Zhang, Ying Zhang, Zhaxisangzhu, X.~X. Zhou, and The
  Tibet AS~γ Collaboration.
\newblock Search for gamma rays above 100 tev from the crab nebula with the
  tibet air shower array and the 100 m2 muon detector.
\newblock {\em The Astrophysical Journal}, 813(2):98, 2015.

\bibitem{ARGO}
B.~Bartoli et~al.
\newblock {Crab Nebula: five-year observation with ARGO-YBJ}.
\newblock {\em Astrophys. J.}, 798(2):119, 2015.

\bibitem{LHAASO}
G.~Di~Sciascio.
\newblock {The LHAASO experiment: from Gamma-Ray Astronomy to Cosmic Rays}.
\newblock {\em Nucl. Part. Phys. Proc.}, 279-281:166--173, 2016.

\bibitem{ALPACA}
M.~Ohnishi et~al.
\newblock {The overview of the ALPACA Experiment}.
\newblock {\em Proceedings of Science: 35th International Cosmic Ray
  Conference}, 437, 2017.

\bibitem{ALTO}
Y.~Becherini et~al.
\newblock {Very-High-Energy gamma-ray astronomy with the ALTO observatory}.
\newblock {\em Proceedings of Science: 35th International Cosmic Ray
  Conference}, 782, 2017.

\bibitem{LATTES}
R~Concei\c{c}\~{a}o et~al.
\newblock {LATTES: a novel detector concept for a gamma-ray experiment in the
  Southern hemisphere}.
\newblock {\em Proceedings of Science: 35th International Cosmic Ray
  Conference}, 784, 2017.

\bibitem{STACEX}
G~Di~Sciascio et~al.
\newblock {Gamma-Ray Astronomy with a Wide Field of View detector operated at
  Extreme Altitude in the Southern Hemisphere.}
\newblock {\em Proceedings of Science: 35th International Cosmic Ray
  Conference}, 781, 2017.

\bibitem{MPI}
H~Schoorlemmer et~al.
\newblock {Baseline Design for a Next Generation Wide-Field-of-View
  Very-High-Energy Gamma Ray Observatory}.
\newblock {\em Proceedings of Science: 35th International Cosmic Ray
  Conference}, 819, 2017.

\bibitem{ref_CORSIKA}
D.~{Heck}, J.~{Knapp}, J.~N. {Capdevielle}, G.~{Schatz}, and T.~{Thouw}.
\newblock {\em {CORSIKA: a Monte Carlo code to simulate extensive air
  showers.}}
\newblock February 1998.

\bibitem{ref_GEANT4}
S.~Agostinelli et~al.
\newblock {GEANT4: A Simulation toolkit}.
\newblock {\em Nucl. Instrum. Meth.}, A506:250--303, 2003.

\bibitem{ref_HAWC_crab_paper}
A.~U. Abeysekara et~al.
\newblock {Observation of the Crab Nebula with the HAWC Gamma-Ray Observatory}.
\newblock {\em Astrophys. J.}, 843(1):39, 2017.

\bibitem{ref_MINUIT}
F.~{James} and M.~{Roos}.
\newblock {Minuit - a system for function minimization and analysis of the
  parameter errors and correlations}.
\newblock {\em Computer Physics Communications}, 10:343--367, December 1975.

\bibitem{hawc_Outriggers1}
V.~{Joshi}.
\newblock {HAWC High Energy Upgrade with a Sparse Array}.
\newblock In {\em European Physical Journal Web of Conferences}, volume 136 of
  {\em European Physical Journal Web of Conferences}, page 03006, March 2017.

\end{thebibliography}

\end{document}